\documentclass[%
 aip,
 amsmath,amssymb,
 reprint,%
floatfix,
nofootinbib
]{revtex4-1}

\usepackage{mathtools}

\usepackage{graphicx}
\usepackage{dcolumn}
\usepackage{bm}

\usepackage[utf8]{inputenc}
\usepackage[T1]{fontenc}
\usepackage{mathptmx}
\usepackage{etoolbox}
\usepackage{url}

\usepackage{multirow}

\usepackage{subcaption}
\DeclareCaptionFormat{groupplotformat}{#1 #3}
\usepackage{xcolor}
\usepackage{tikz}
\usepackage{pgf,pgfplots}
\usetikzlibrary{
arrows,%
backgrounds,%
calc,%
decorations.text,%
decorations.pathreplacing,%
external,%
fadings,%
fit,%
intersections,%
matrix,%
shadings,%
shadows,%
shapes,%
shapes.multipart,%
spy,%
patterns,%
plotmarks,%
through}
\pgfplotsset{compat=1.18} 
\usepgfplotslibrary{groupplots,fillbetween}

\pgfplotsset{
    tuft/.style = {
        width=\columnwidth,
        clip=false,
        axis lines*=left,
        mark size=2,
        axis line shift=15pt,
        tick style={thick,black},
        major tick length=0.15cm,
        x tick label style={rotate=-60,anchor=west}
    }
}

\definecolor{lab6color}{RGB}{144,59,209}
\definecolor{cathodecolor}{RGB}{100,100,100}
\definecolor{plasmacolor}{RGB}{242,170,250}

\makeatletter
\def\@email#1#2{%
 \endgroup
 \patchcmd{\titleblock@produce}
  {\frontmatter@RRAPformat}
  {\frontmatter@RRAPformat{\produce@RRAP{*#1\href{mailto:#2}{#2}}}\frontmatter@RRAPformat}
  {}{}
}%
\makeatother

\begin{document}

\def\mytitle{Rarefied {xenon} flow in orificed hollow cathodes}
\title[\mytitle]{\mytitle}
\author{Pierre-Yves C. R. Taunay\textsuperscript{*}}
\affiliation{Aerospace Engineering Department, United States Naval Academy, Annapolis, MD 21402, USA}
\email{taunay@usna.edu}
\author{Willca Villafana}%
\author{Sangeeta P. Vinoth}%
\author{Igor Kaganovich}
\author{Andrei Khodak}%
\affiliation{ 
Princeton Plasma Physics Laboratory, Princeton, NJ 08543, USA
}

\date{\today}

\begin{abstract}
A parametric study is conducted to quantify the effect of the keeper electrode geometry on the {xenon} neutral flow quantities 
within orificed hollow cathodes, prior to ignition. The keeper impinges directly on the flow out of the cathode orifice 
and its geometry influences the product between the pressure in the orifice-keeper region and the cathode-to-keeper distance. 
A representative cathode is simulated using the Direct Simulation Monte Carlo method. The numerical model is first 
validated with computational results from the literature. A parametric study is then conducted. Parameters include the 
cathode pressure-diameter in the range of 1--5 Torr-cm and the following geometric ratios (and ranges): cathode orifice-to-inner radii (0.1--0.7), 
keeper orifice-to-cathode orifice radii (1--5), and keeper distance-to-cathode-orifice diameter (0.5--10). 
It is found that, if both keeper and cathode have identical orifice radii, the flow remains subsonic in the orifice-to-keeper region. 
In most cases, however, the flow becomes underexpanded and supersonic, and the static pressure within the orifice-to-keeper 
region is, on average, 4\% that of the upstream pressure value. The orifice-keeper region pressure increases with either a 
decrease in the keeper orifice diameter or an increase in the distance between cathode and keeper, in agreement with literature data. 
Both trends are explained through conservation laws. A statistical study of numerical results reveals that the ratio of 
ignition-to-nominal mass flow rates has a most probable value of 50, which suggests that heaterless cathode ignition at 
a minimum DC voltage may be achieved by increasing the input mass flow rate by a factor of 50.
\end{abstract}

\maketitle

\onecolumngrid
\noindent
\fbox{
\parbox{0.97\textwidth}{
\textbf{
The following article has been submitted to the Journal of Applied Physics. \\
\noindent
{Copyright (2024) Pierre-Yves C. R. Taunay, Willca Villafana, Sangeeta P. Vinoth, Igor Kaganovich, Andrei Khodak. This article is distributed under a Creative Commons Attribution-NonCommercial 4.0 International (CC BY-NC) License. \url{https://creativecommons.org/licenses/by-nc/4.0/}}
}
}
}
\vspace{20pt}
\twocolumngrid

\section{Introduction}
Thermionic, orificed hollow cathodes are sources of electrons that are widely used in 
space electric propulsion,\cite{Goebel2007,Mikellides2008-1,Mikellides2008-2,Lev2019,Lev2019-2} where they deliver
electrons for the main plasma discharge and to neutralize the
ion exhaust beam.
These cathodes consist of a tube 
that is capped by an orifice plate
and that contains a thermionic emitter cylinder (\textit{e.g.}, barium-oxide). 
A keeper electrode, typically with a single orifice (or multiple orifices\cite{Georgin2021}), faces the main cathode orifice plate downstream
of the cathode tube. 
The keeper is used to establish the initial plasma discharge
and to protect the cathode from high-energy ions that originate in the plume
of the cathode and stream back towards the cathode.

A heater may be used to bring the emitter to electron emission temperatures prior
to the start of the thermionic discharge. In this approach, the emitter is
heated to high temperatures ($>1,200^\circ$C), a neutral gas (\textit{e.g.}, xenon)
is inserted in the cathode tube, and a discharge is established between the 
keeper and cathode.
Alternatively, a DC, Paschen-like discharge may be established between the cathode and the keeper to start the cathode.
Gas is inserted in the cathode tube, and a glow discharge produced by a high-voltage ($>100$~V) DC source  
effectively heats the cathode orifice and emitter.
This approach is favored for low-to-medium amperage cathodes and circumvents
the use of a heater: the heater and its associated power supply are oftentimes considered to be a single point-of-failure for
space electric propulsion systems.
The accurate placement and sizing of the keeper is critical for a reliable DC discharge: 
an optimal value of the pressure-distance product, 
$P_{ko} D_{ko}$,
results in a minimum in the applied voltage required for the glow discharge,
where $P_{ko}$ is the pressure of the gas in the orifice-keeper region, and
$D_{ko}$ is the (axial) distance between the keeper and the cathode orifice plate.
In the context of spacecraft electric propulsion, minimizing the applied voltage is desirable as it 
reduces the amount of on-board electric insulation required when using high-voltage power supplies.\cite{Lev2019}
Although the keeper-orifice geometry is not identical to that of a Paschen discharge 
(\textit{e.g.}, the cathode orifice and keeper are not truly planar surfaces), 
the minimum pressure-distance product is typically $\sim$~{1--5~Torr-cm}.\cite{Lev2019}

An orificed hollow cathode that operates without a plasma discharge is similar to a supersonic neutral beam source:
the cathode consists of a high-pressure vessel 
(relative to the exhaust vessel, $P_c d_c\sim$1--10~Torr-cm, where $P_c$ is the cathode upstream pressure and $d_c$ 
its internal diameter) that expands into a vacuum background ($P_0\sim$1~$\mu$Torr).
Because both the mean free path in the plume and the ratio of cathode-to-background pressure 
are large ($\geq 10^3$), shocks may be diffuse and the typical structure 
observed in supersonic neutral beams, including the ``zone of silence,''
may not appear.\cite{Campargue1984}
The presence of a keeper, similar to a ``skimmer'' in supersonic
neutral beams, affects the structure of the flow 
downstream of the orifice. 

Beyond the orifice and keeper, the neutral gas density has a direct effect on the
generation of high-energy ions when the cathode operates with a plasma.
Locally increasing the neutral density by injecting neutral gas in front of the keeper
yields a decrease in energy of the high-energy ions generated in the plume.\cite{Chu2013,Goebel2014}
The neutral density field in this region is a critical quantity to accurately
represent collisions between neutral and charged particles in numerical approaches
such as particle-in-cell (PIC).

It is challenging to model the gas flow through the cathode to
obtain a reliable value of $P_{ko}$ and study the flow structure beyond the
orifice and keeper: 
the flow is compressible and rarefied, 
with Knudsen numbers, $\textrm{Kn}$, above $0.1$.\cite{Taunay2022}
For monatomic gases, the Mach number might be greater than one at the orifice exit,
\cite{Lilly2004,Varoutis2008} indicating that
a transition from subsonic to supersonic flow also exists within the orifice.
Theoretical descriptions of the neutral flow through cathodes and/or tubes
include isentropic,\cite{Domonkos1999} Poiseuille,\cite{Goebel2008,Becatti2021},
and free-molecular flow\cite{Cai2007} models, as well as empirical approaches.\cite{Santeler1986}
Fundamental assumptions on which the isentropic or Poiseuille flow descriptions
rely are invalid in the flow regime in which cathodes operate.\cite{Taunay2022}
While the empirical approach suggested by Santeler\cite{Santeler1986} provides transitional-flow corrections
to the cathode pressure, it cannot describe the flow field in the orifice and 
in the plume.
The collisionless approach of Cai and Boyd\cite{Cai2007} is valid in the plume of
the cathode, where the flow is nearly free-molecular. However, it cannot describe
the transitional flow in the orifice and is unable to capture the
effect of the keeper on the flow past the orifice.

Cathode numerical studies that feature both charged and neutral particles 
include
continuum approaches
and/or hybrid-continuum approaches,~\cite{Mikellides2005,Mikellides2006,Mikellides2009,Mikellides2014,Sary2017-I} 
wherein the neutrals are in the continuum regime within the cathode tube and in the molecular flow regime in the plume, 
as well as
Direct Boltzman solvers\cite{Vazsonyi2020,Vazsonyi2020-2,Vazsonyi2021}
and PIC approaches.\cite{Cao2019}
The assumption of a continuum and free-molecular flow are valid within the dense upstream emitter region
of the cathode and in the cathode plume, respectively.
However, it is invalid both in the orifice and in the orifice-keeper region: the flow is transitional 
in both regions. 
Direct Boltzmann solvers, unlike PIC methods, are not subject to statistical noise. 
However, they are typically limited to the plume
region because of their computational expense when applied to the dense and collisional internal region of the cathode. 
The study of Cao~\textit{et al.}\cite{Cao2019} features both a PIC approach for the charged particles
and the Direct Simulation Monte Carlo (DSMC) technique to solve for the neutral quantities (\textit{e.g.}, density). 
However, Cao~\textit{et al.}\cite{Cao2019} did not include the keeper electrode as part
of the simulation domain and focused only on the generation
of a neutral flow field that can be used as input to a Particle-In-Cell (PIC) 
solver for a single test case. 

In general, past DSMC studies solely dedicated to hollow cathodes have been limited.
While not strictly targeted at hollow cathodes, 
the study of Varoutis~\textit{et al.}\cite{Varoutis2008} includes a wide range of
orifice aspect ratio (\textit{i.e.}, ratio of cathode orifice length to orifice diameter) and is broadly applicable to 
keeper-less cathodes. 
Good agreement with experimental data 
has been reported for keeper-less cathodes.\cite{Nikrant2021}
Daykin-Iliopoulos~\textit{et al.}\cite{Daykin2015} focused on the 
keeper-orifice region in enclosed-keeper cathodes and 
considered a 
wide range of parameters that consist of cathode-to-keeper distance,
keeper-to-orifice diameter ratios, and orifice-to-insert diameter ratios. 
Daykin-Iliopoulos~\textit{et al.} observed that the pressure within the
keeper-orifice region increases with increasing electrode separation and
decreasing keeper diameter. 
However, this study did not provide
a physical explanation of the observed trends.

In this work, we choose to revisit the study of Daykin-Iliopoulous~\textit{et al.} and
explore, in depth, the effect of the keeper electrode on the 
fluid flow within orificed hollow cathodes that feature an enclosed keeper. 
Using a DSMC approach, we seek to 
identify which geometrical parameters
most significantly impact the pressure-distance product
in the orifice-keeper region of cathodes, provide broadly applicable
design rules, and explain, physically, the observed trends.
We also compare DSMC results to that of a fluid solver in select cases to 
understand the limitations of continuum approaches applied to rarefied gas flows.
We first describe our numerical approach and validate our model based on 
reported literature data in Section II.  
We then perform a parameterized study based on set ranges of non-dimensional parameters that are reported in Section III.
Scaled results are presented and discussed in Section IV.

\vspace{20pt}
\section{Implementation}

\subsection{Direct Simulation Monte Carlo}

The DSMC method is a numerical technique introduced by Bird\cite{Bird1994} that may be used to
simulate gases that are in the non-continuum regime. 
The DSMC method aims to solve numerically the Boltzmann equation with a probabilistic, particle-based approach: 
each numerical ``particle'' represents a large number of physical particles, and 
particle-particle collisions are handled with a probabilistic representation.
The method has been applied across a variety of engineering and scientific domains, 
such as the aerodynamics of the Space Shuttle Orbiter\cite{Rault1994} and 
the fluid flow in micro and nano devices.\cite{Fang2002}

\subsection{Software}

We use the DSMC code ``Stochastic PArallel Rarefied-gas Time-accurate Analyzer'' (SPARTA)\cite{Plimpton2019,SPARTA} for our simulations.
SPARTA is a scalable and open-source DSMC code, parallelized using Message-Passing Interface (MPI).
The software has been extensively validated\cite{Gallis2015,Gallis2016,Gu2019} and is commonly used
in the DSMC community.

DSMC results are validated with a separate fluid solver, ANSYS-CFX.
ANSYS-CFX is based on the continuum approach (\textit{i.e.}, mass, momentum, and energy equations are solved).
ANSYS-CFX uses an extended upstream region which differs from that of the DSMC analysis. 
If both software are applied to the same region, the results would be in closer quantitative agreement.
However, for simplicity and computational expediency, the DSMC code is only applied to a truncated domain.
This is discussed in Section III.B.

\subsection{Numerical model}
\subsubsection{Physical domain and boundary conditions}

\begin{figure*}[h]
    \centering
    \includegraphics[scale=1.1]{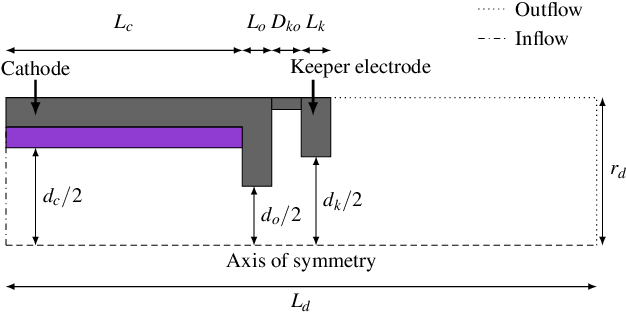}
    \caption{Cathode computational representation.}
    \label{fig:physical-domain}
\end{figure*}

Figure~\ref{fig:physical-domain} shows the physical domain representing a cathode equipped with an enclosed keeper,
along with the relevant physical parameters. 
While some cathodes\cite{Polk1999,Becatti2021} feature a chamfered orifice, we choose not to include it for simplicity.

\paragraph{Wall boundary condition}
Particles may interact with the walls through specular or diffuse collisions.
Specular reflections result in a sign change for the component of the particle velocity that is
perpendicular to the wall. The component of the particle velocity that is parallel to the wall
remains unchanged, which results in all odd-order moments perpendicular to the wall vanishing.
The particles cannot transfer momentum nor heat --- the flow is effectively inviscid.

We choose to apply diffuse boundary conditions throughout {with a choice of
a constant accommodation coefficient for all solid surfaces.}
{
While the value of the accommodation coefficient may change appreciably depending
on the surface material and roughness,\cite{Agrawal2008,Selden2009} a sensitivity study
of the solution on the accommodation coefficient is beyond the scope of this article:
we do not consider the influence of gas-surface interactions on our solution
and set the accommodation coefficient to a value of 0.8.
While this value is 13\% lower than the recommended value from Agrawal and Prabhu,\cite{Agrawal2008}
it is within the range of other reported data for xenon interaction with solid walls\cite{Selden2009}
and, therefore, represents a conservative value of the accommodation coefficient.}
Other studies\cite{Sharipov2004} have shown that the flow rate through an orifice is 
not sensitive to the value of the accommodation coefficient
for vessels separated by a thin orifice.

\paragraph{Outflow boundary condition}
Particles that reach the edge of the domain where an outflow condition is imposed simply vanish. 
They are not re-injected into the domain.

\paragraph{Inflow boundary condition}
The upstream surface features a subsonic inlet with a set pressure and temperature
{, as opposed to a set mass flow rate.}
A varying number of particles is injected to enforce the desired pressure and temperature.
The total number of injected particles accounts for domain particles that exit the simulation domain 
through the upstream surface.
{The mass flow rate for the numerical experiments, $\dot{m}_\textrm{exp}$, 
may be retrieved by computing the flux of particles
that cross a plane intersecting the cathode channel.
The mass flow rate typically depends on the orifice geometry and may differ from the theoretical
values that can be computed using a typical isentropic flow approach, $\dot{m}_{th}$.
We define the discharge coefficient, $C_d$,
as the ratio of the experimental-to-theoretical flow rates:
\begin{equation}
    C_d = \dfrac{\dot{m}_\textrm{exp}}{\dot{m}_{th}}.
\end{equation}
}

\onecolumngrid

\twocolumngrid

\subsubsection{Collision model}
Particle-particle collisions are represented by the variable soft sphere (VSS) model, 
{
in which the relationship between the collision cross-section
and the collision energy is represented by a power law.
For this model, the collision deflection angle is 
$\chi = 2 \arccos( (b/d)^{\left(1/\alpha\right)})$,
where $b$ is the impact parameter.
Consequently, the gas viscosity depends on the temperature through a power law: $\mu \propto T^\omega$.
The model depends on four parameters:
\begin{itemize}
    \item $T_\textrm{ref}$: a reference temperature,
    \item $d_\textrm{ref}$: the particle diameter for the collision cross-section at the reference temperature, $T_\textrm{ref}$,
    \item $\omega$: the temperature dependence of the viscosity, and
    \item $\alpha$: the scattering parameter. 
\end{itemize}
}
The VSS parameters are shown in Table~\ref{tbl:vss} for xenon.
\begin{table}[ht!]
\caption{\label{tbl:vss}VSS parameters for xenon (Ref.~\onlinecite{Bird1994}, pp.~408,410).
{
 $T_\textrm{ref}$, $d_\textrm{ref}$, $\omega$ and $\alpha$ are the reference temperature,
the particle diameter for the collision cross section at the reference temperature ($T_\textrm{ref}$),
the temperature dependence of the viscosity, and
the scattering parameter, respectively.}
}
\begin{ruledtabular}
\begin{tabular}{ccccc}
Species & $d_\textrm{ref}$ (m) & $\omega$ & $T_\textrm{ref}$ (K) & $\alpha$ \\
Xenon & $5.65\times 10^{-10}$ & 0.85 & 273.15 & 1.44
\end{tabular}
\end{ruledtabular}
\end{table}

\subsubsection{Discretization}
\label{sec:discretization}

\begin{figure}[ht!]
    \centering
    \includegraphics{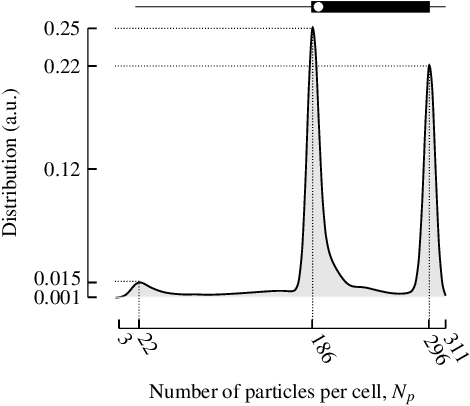}
    \caption{Illustrative distribution and whisker plot of the total number of simulated particles per cell.}
    \label{fig:example-distribution}
\end{figure}

The time step, $\Delta t$, cell size, $\Delta x$ and $\Delta y$, and particle weight 
(\textit{i.e.},
the ratio of real to simulated particles), $f_n$,
must be fine enough to accurately represent 
collisional processes in a DSMC calculation.
The collision mean free path, $\lambda$, collision time scale, $t_c$, and the particle transit time, $t_p$,
shall all be accurately resolved, even for cells with the highest densities and particle velocities:
\begin{itemize}
    \item $\Delta x \approx \Delta y \approx \lambda/N_\lambda$, and
    \item $\Delta t < \textrm{min}\left(t_c/N_t,t_p/N_t\right)$,
\end{itemize}
where $2 \leq N_\lambda, N_t \leq 3$.
The particle weight shall be chosen such that there is a statistically significant number of particles in 
most simulation cells, even for those with the lowest density.
In practice, this means that there should be \textit{at least} 20 particles per cell.
The total number of particles per cell is verified after a given simulation is complete.
Figure~\ref{fig:example-distribution}
shows an example distribution of the number of
particles per cell for a simulation with a locally refined grid. 
While it is clearly multimodal due to the large variations in density between 
the cathode channel and the plume, the number of particles per cell remains
statistically significant in most cells.
For this particular example, the 5\% and 95\% bounds of the distribution are 53 and 297 particles
per cell, respectively, with the most probable peaks
located at 22, 186, and 296 particles per cell. 

\paragraph{Space discretization}
The number of cells in the $j$-th direction is
\begin{equation}
    N_j \approx \dfrac{L_j}{\lambda/N_\lambda},
\end{equation}
where $L_j$ is the domain length in that direction and $\lambda$
is the collision mean free path of neutral atoms.
The total number of cells in a given direction is rounded to the nearest power of two. 
The mean free path for the VSS model, $\lambda$, is given by
\begin{equation}
\label{eqn:vss-mfp}
    \lambda = \dfrac{1}{\sqrt{2} n_n \sigma_T},
\end{equation}
where $n_n$ is the local neutral density and $\sigma_T$ is the total collision cross section. 
The cross section for the VSS model is given by\cite{Bird1994} (pp.~40--42):
\begin{equation}
   \sigma_T = \pi d_\textrm{ref}^2 \left(\dfrac{T_\textrm{ref}}{T}\right)^{\omega-1/2}, 
\end{equation}
where $T$ is the local temperature.

\paragraph{Time step}
The inter-particle collision and transit time scales are given by
\begin{gather}
    t_c = \dfrac{1}{n_n \sigma \bar{v}}, \textrm{and}\\
    t_p = \dfrac{\textrm{min}\left(\Delta x, \Delta y\right)}{\bar{v}},
\end{gather}
where $\bar{v}$ is the most probable velocity. Assuming that particles follow a Maxwellian distribution,
\begin{equation}
    \bar{v} = \sqrt{\dfrac{2 k_B T_n}{M}},
\end{equation}
where $k_B = 1.38\times10^{-23}$~J/(kg$\cdot$K) is the Boltzmann constant,
$T_n$ is the neutral gas temperature, and $M$ the mass of a single gas atom.

\begin{figure}[ht!]
    \centering
    \includegraphics{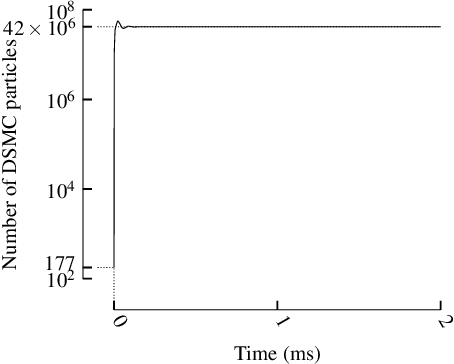}
    \caption{Typical evolution of the total number of particles.}
    \label{fig:convergence-example}
\end{figure}

\clearpage

\onecolumngrid

\begin{figure*}[h!]
    \centering
    \includegraphics[width=6.0in]{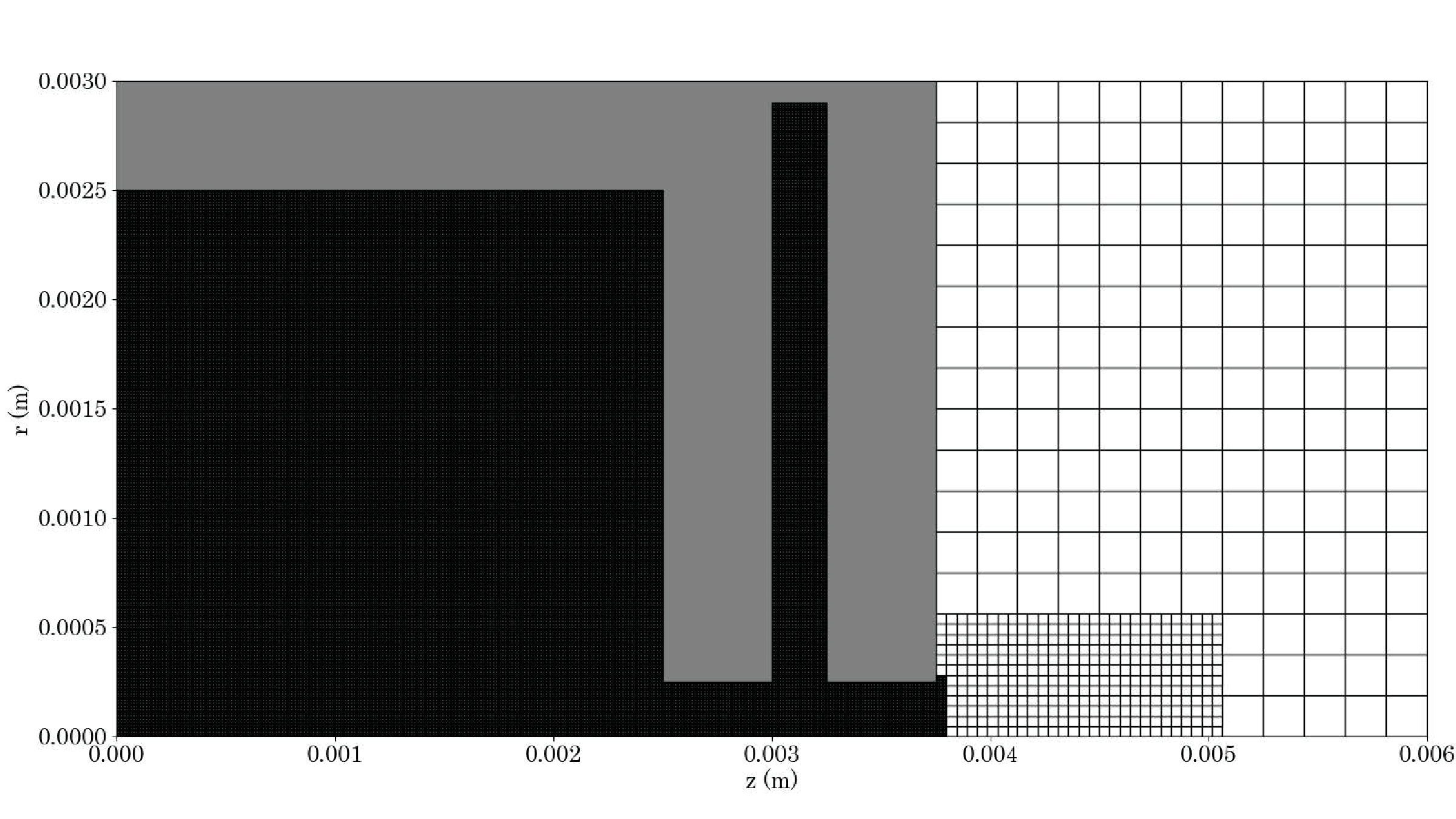}
    \caption{Illustrative adapted grid for the case where $d_k/d_o = 1$ and $D_{ko}/d_o = 0.5$. The cathode surface is indicated in gray.
    The internal grid is 32$\times$ more refined than the plume grid.}
    \label{fig:adapted-grid}
\end{figure*}

\twocolumngrid

\paragraph{Particle weight}
In axisymmetric coordinates, the volume of a given cell is given by
\begin{equation}
    V_j = \pi \left( (j\Delta y)^2 - (j-1)\Delta y)^2\right)\Delta x,
\end{equation}
where $j$ is the index of the cell along the $y$-axis.
The corresponding number of real and simulated particles is $V_c n_n$ and ${V_c n_n/f_n}$
respectively.
The particle weight is chosen such that the cell with the smallest volume 
(\textit{i.e.}, the center cell in axisymmetric coordinates) has at least 20 simulated particles,
even in the lowest-density regions. 
Because the volume of the axisymmetric cell increases with radius, cells on the outermost
boundary of the domain will contain a larger number of simulated particles.
To ensure that each cell contains an adequately balanced number of particles, we 
assign a radius-based weight to the cells, which internally rescales
$f_n$ by the local cell volume in SPARTA. 

\subsubsection{Convergence and statistics}
The domain is initialized with a non-zero number density that is computed based on a vacuum condition of 1~$\mu$Torr and a gas temperature of 300~K (\textit{i.e.}, $n_n=3.2\times 10^{16}$~m\textsuperscript{-3}).
Particles are drawn from a Maxwellian distribution with zero streaming velocity
and the upstream temperature. 
{After initialization, additional particles are injected continuously at the inflow boundary described in Section II.C.1.c at a set temperature and pressure.}
{Injected particles travel throughout the domain and interact with the different
boundary conditions. They are removed from the domain if they reach the outflow (or the inflow)
boundary.}

The simulation is considered to have converged once the total number of particles
remains approximately constant. 
Figure~\ref{fig:convergence-example} shows 
the typical evolution of the total number of (simulated) particles as a function of time.
Because the initial density in the cathode is much lower than the steady-state one (approximately two-to-three
orders of magnitude) the total number of particles increases until it reaches steady-state.

Relevant physical quantities {(\textit{e.g.}, the gas temperature or the mass flow rate)} are computed once steady-state is reached. 
Results are averaged over 10,000
time steps ($\sim$~0.1~ms) to decrease
the statistical noise inherent to particle simulations.

\subsection{Multi-level grid and load rebalance}
The time-to-solution is reduced by \textit{(i)} decreasing the overall number of cells
within the domain with a multi-level grid and \textit{(ii)} rebalancing the workload across
multiple processors as particles appear within or vanish from the domain.
The grid is refined within the cathode tube and keeper regions. A coarse grid is
sufficient for the plume region. An illustrative, adapted grid is shown in
Figure~\ref{fig:adapted-grid}. 
Care is taken to not extensively refine the low-density 
region past the corner of the keeper exit.
Because the particle weight, $f_n$, is set globally for the simulation, 
a refined grid in this region would result in a low number of DSMC
particles per computational cell.

Because the number of DSMC particles increases significantly within the first 
10,000 steps of the simulation (see, \textit{e.g.}, Figure~\ref{fig:convergence-example}),
we rebalance the particle load across processors every 100 steps during that time period. Periodic rebalancing is applied throughout the simulation thereafter. 

\clearpage

\onecolumngrid

\begin{figure*}[h]
    \centering
    \includegraphics{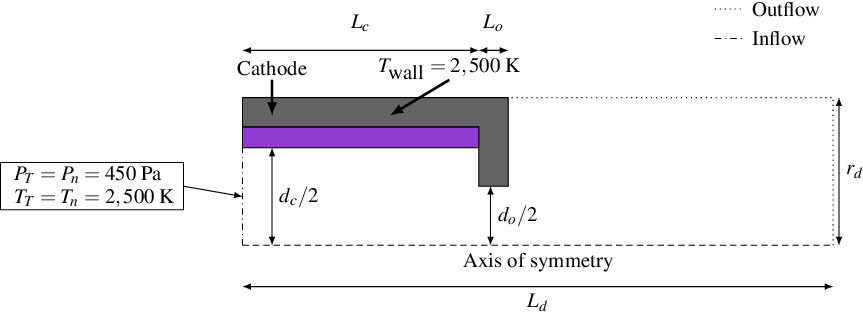}
    \caption{Cathode computational representation for the numerical validation test case.}
    \label{fig:cao-physical-domain}
\end{figure*}

\twocolumngrid

\subsection{Numerical validation}

We validate the DSMC model with the test case described by Cao~\textit{et al.},\cite{Cao2019} which features a computational domain without a keeper electrode. 
{The computational domain for this test case is shown in Figure~\ref{fig:cao-physical-domain}.}
A fully coupled PIC/DSMC approach wherein both charged
and neutral particles are simulated is beyond the scope of this article:
we cannot capture plasma-neutral effects 
such as the increase in neutral particles temperature near the orifice due to the frequent charge-exchange collisions 
with the plasma ions.

The parameters describing the computational domain of the validation case are shown
in Table~\ref{tbl:cao-main-geometry}. 
{Using the input parameters provided by Cao~\textit{et al.}\cite{Cao2019} 
for the upstream pressure and density 
(450~Pa and 1.3$\times$10\textsuperscript{22}~m\textsuperscript{-3}, respectively)
and the perfect gas law,} the temperature of the injected gas is {found} to be 2,500~K. 
Because we are considering a heavy, noble gas that is monatomic at high temperatures and
low densities, the thermal de Broglie wavelength ($\mathcal{O}\left(10^{-12}\,\textrm{m}\right)$)
is much smaller than the size of the domain ($\mathcal{O}\left(10^{-3}\,\textrm{m}\right)$): 
Boltzmann statistics still apply,\cite{McQuarrie} and the perfect gas law is applicable.
Unlike Cao~\textit{et al.}, we do not rescale the heavy particle mass or domain size.
For the set input upstream conditions, this results in an overestimation of the mass flow rate within the cathode.
However, quantities such as the (static) pressure and the number density can be directly compared. 

The data reported by Cao~\textit{et al.}\cite{Cao2019} indicate that the neutral gas temperature is kept constant
throughout the domain. We apply a wall temperature equal to that of the
injected particles (\textit{i.e.}, $\approx$2,500~K) to replicate Cao~\textit{et al.}'s results within the internal
cathode region.
However, there is no mechanism by which we can enforce a constant static temperature 
in the entire domain.
\begin{table}[ht!]
\caption{\label{tbl:cao-main-geometry}Physical domain for the test case presented in Cao~\textit{et al.}\cite{Cao2019}}
\begin{ruledtabular}
\begin{tabular}{cccc}
Parameter & Unit & Symbol & Value \\
Inner channel diameter & mm  & $d_c$ & 5\\
Orifice diameter & mm & $d_o$ & 2\\
Orifice length & mm & $L_o$ & 1\\
Domain length & mm & $L_d$ & 16\\
Domain radius & mm & $r_d$ & 4\\
Upstream pressure & Pa & $P_n$ & 450\\ 
Upstream density & m\textsuperscript{-3} & $n_n$ & $1.3\times10^{22}$ \\
Upstream temperature & K & $T_n$ & 2,500\\
Neutral species mass & amu & $M_a$ & 131.293
\end{tabular}
\end{ruledtabular}
\end{table}

Under the assumption that the flow is choked
and reaches Mach 1 at the orifice exit, 
the static pressure at the exit plane of the orifice, $P_\textrm{exit}$,
may be estimated with either an isentropic flow or a Fanno flow model.
In both cases, the static pressure is given by
\begin{equation}
    P_\textrm{exit} = P_\textrm{T,exit} 
    \left(\dfrac{2}{\gamma+1}\right)^{\gamma/\left(\gamma-1\right)},
\label{eqn:Pexit_from_Ptexit}
\end{equation}
where $P_\textrm{T,exit}$ is the total pressure at the exit. 
For an isentropic flow, the total pressure remains constant, and $P_\textrm{T,exit} = P_T$, where $P_T$ is the provided upstream pressure (here, $P_T=450$~Pa).
Alternatively, because the Reynolds number in the orifice is low ($\textrm{Re} \sim 10$--$10^2$), a Fanno flow approach may be considered. 
For a Fanno flow, the ratio of total pressures across the orifice is\cite{White}
\begin{equation}
\dfrac{P_\textrm{T,in}}{P_\textrm{T,exit}} = \dfrac{1}{M_{in}} \left[ \dfrac{2+\left(\gamma-1\right)M_{in}^2}{\gamma+1}\right]^{\dfrac{\gamma+1}{2\left(\gamma-1\right)}},
\end{equation}
where the subscripts ${in}$ and ${exit}$ denote the orifice entrance and exit quantities, respectively.
For this particular validation case, the entrance Mach number is found to be $M_{in} \approx 0.5$ from the computational results.
Under the assumption that the flow is isentropic in the cathode emitter region, $P_\textrm{T,in} = P_\textrm{T}$, and
Equation~\ref{eqn:Pexit_from_Ptexit} may then be used to obtain $P_\textrm{exit}/P_T$.

A Poiseuille flow approach is often used to estimate the pressure within the cathode.\cite{Goebel2008,Mizrahi2012,Becatti2021,Becatti2022,Poli2023}
We use the model from Goebel and Katz\cite{Goebel2008} with the xenon viscosity from Stiel and Thodos\cite{Stiel1961}. 
The mass flow rate is computed from the simulation, as described in Section II.C.1.c, and is found
to be 1.5~mg/s (15.3~sccm).

\begin{figure}[ht!]
    \centering
    \includegraphics{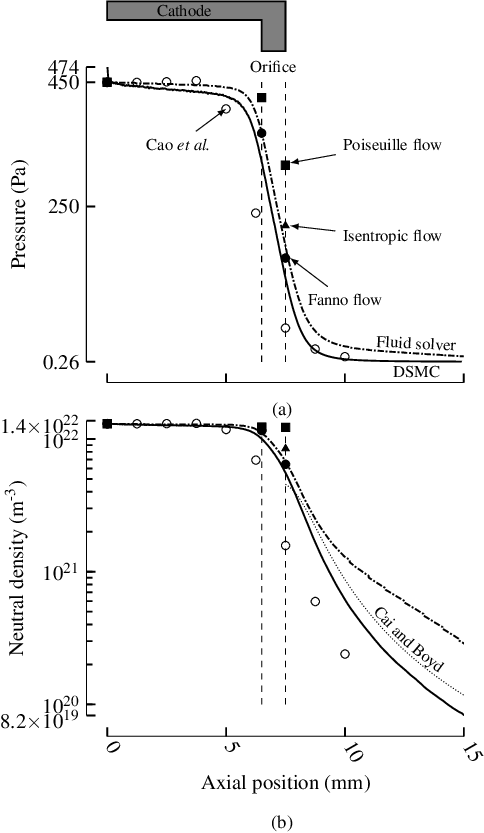}
    \caption{On-axis (a) neutral pressure and (b) density for Cao~\textit{et al.}'s test case. All flow models start with the same value at $z=0$~mm. The isentropic and Fanno flow
    models are assumed to have the same static quantities at the orifice entrance. 
    }
    \label{fig:cao-results}
\end{figure}

Figure~\ref{fig:cao-results} compares the pressure and the neutral density as obtained with
our theoretical and numerical approaches to that of Cao~\textit{et al.}
We also compare our results to the collisionless free-jet solution
obtained by Cai and Boyd\cite{Cai2007}
and to the results of the fluid solver ANSYS-CFX.
The average orifice exit (static) temperature, gas velocity, and density scale required for
the analytical model of Cai and Boyd\cite{Cai2007} are derived from our simulation.
They are equal to 1,823~K, 447~m/s, and $5.0\times 10^{21}$~m\textsuperscript{-3}, respectively.
{We estimate the average orifice gas temperature for the Poiseuille flow 
approach by assuming that the orifice is choked (\textit{i.e.}, the Mach number in the orifice is equal to one).
The orifice static temperature used for the Poiseuille flow model is found to be equal to 1,873~K.}
Reasonable agreement is obtained for all quantities for all numerical approaches.
Because the flow is transitional in the orifice and near-orifice region (\textit{i.e.}, Kn$\sim$0.1--1), our numerical results differ from both 
the collisionless solution and the fluid solution.
The DSMC and collisionless approaches yield the same 
density decay in the far field ($z\geq 12$~mm) where the Knudsen number increases.
However, the fluid solver is unable to resolve this density decay. 
The Fanno flow model of the orifice yields the most accurate theoretical value. 
The Poiseuille flow approach overestimates the pressure  
{by 32\% and 134\% at the orifice entrance and exit, respectively.}
As discussed elsewhere,\cite{Taunay2022} this approach should not be used to estimate the pressure within cathodes as it is inapplicable to the flow regime in which
cathodes operate: most of the assumptions required by this flow model are invalid in cathodes. For example, the flow is not fully developed in the orifice and compressibility
effects are significant since the flow becomes sonic at the orifice exit.

\begin{figure*}
\subfloat[Neutral density contour]{\label{fig:cao-contour-nn}
\centering
\begin{tikzpicture}
\node at (0,0) {\includegraphics[width=6.5in]{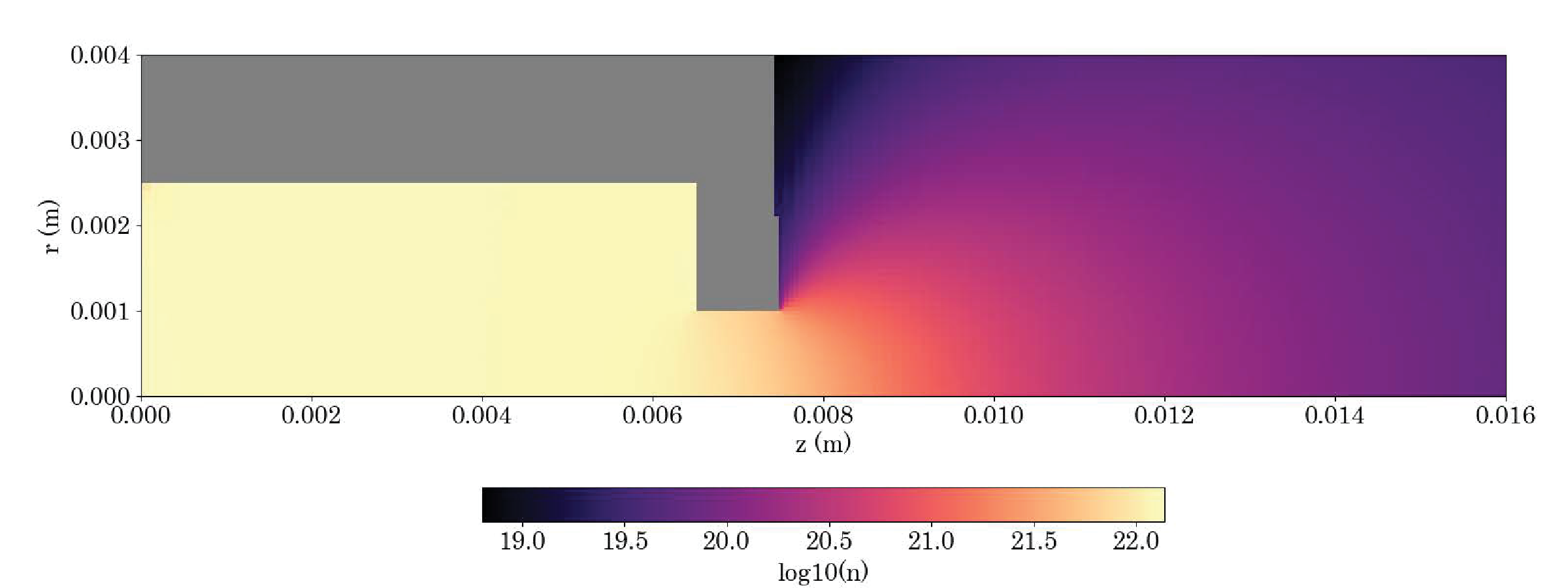}};
\node[fill=white] at (0.5,-3.0) {$\log_{10}\left(n_n\right)$};
\end{tikzpicture}
}

\subfloat[Knudsen number contour] {\label{fig:cao-contour-Kn}
\centering
\begin{tikzpicture}
\node at (0,0) {\includegraphics[width=6.5in]{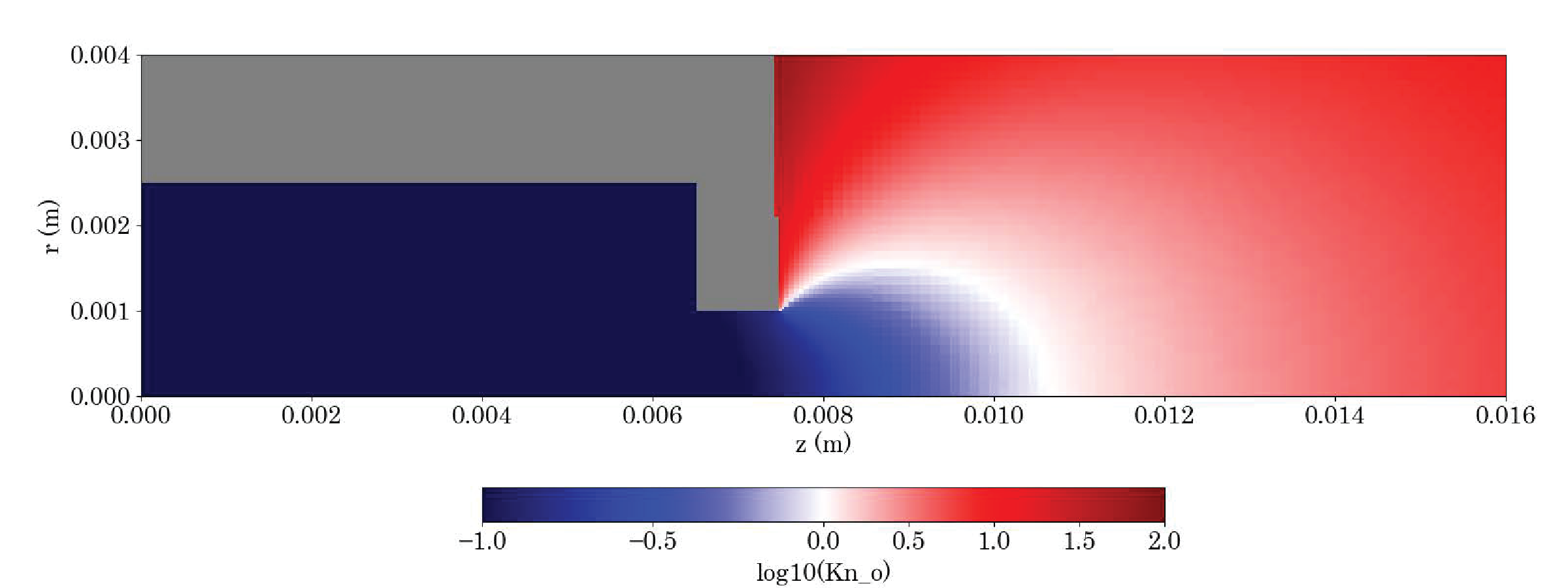}};
\node[fill=white] at (0.5,-3.0) {$\log_{10}\left(\textrm{Kn}_o\right)$};
\end{tikzpicture}
}

\caption{Contours of (a) neutral density and (b) Knudsen number (normalized by the orifice diameter) for Cao~\textit{et al.}'s test case {as obtained with our DSMC approach.}}
\label{fig:cao-contours}
\end{figure*}

Figure~\ref{fig:cao-contours}(a) shows the neutral density contour for this test case.
The Knudsen number (defined here with the characteristic length equal to the orifice diameter, $L=d_o$) is shown in Figure~\ref{fig:cao-contours}(b) and illustrates
that, for this particular test case, the cathode operates in the transitional ($\textrm{Kn}\sim 0.1$) to 
free molecular ($\textrm{Kn} \geq 10$) regimes.

A keeper electrode located at $\sim1$~mm downstream of the orifice would
impinge on the flow and affect local, static quantities. 
We explore the effect of a keeper electrode placement and size in the next sections.

\section{Parametric study}

\subsection{Scope}
\begin{figure}[ht!]
    \centering
    \includegraphics{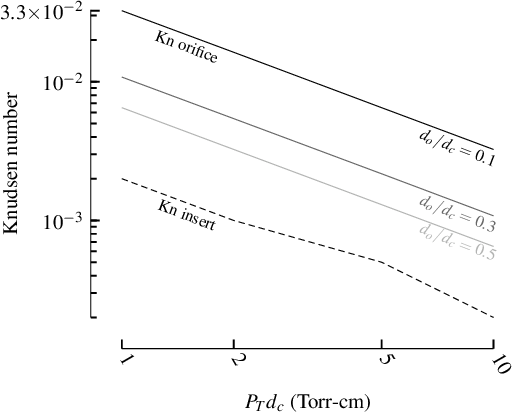}
    \caption{Range of insert and orifice Knudsen numbers for the representative cathode.}
    \label{fig:physical-parameters-range}
\end{figure}

We now consider a representative 
hollow cathode with an inner diameter of 0.5~cm
and investigate the influence of the keeper electrode on the neutral flow. 
Most cathodes operate in a limited range of internal pressure-inner diameter products of 1--10~Torr-cm.\cite{TaunayPSST2022-2} 
During operation, 
the neutral gas is heated by frequent charge-exchange collisions 
and reaches temperatures that are estimated to be 2,000--4,000~K.\cite{Goebel2008}
Because this work is concerned with the neutral flow \textit{prior} to ignition, the 
upstream temperature of the neutral gas is, instead, set to a value of $T_T=300$~K. 
The neutral density within the cathode and orifice is, therefore, higher than that of
the high-temperature cases, and the Knudsen number is, correspondingly, lower.
The Knudsen number in the cathode may be evaluated by assuming that the cathode orifice is
choked, with a Mach number of $1$ within the orifice. The corresponding static
temperature in the orifice is 225~K.
Figure~\ref{fig:physical-parameters-range} shows the corresponding range of insert and orifice
Knudsen numbers for this representative cathode, 
computed with Equation~\ref{eqn:vss-mfp} for the mean free path and using 
the cathode insert or orifice diameter as the relevant length scale. 
In most cases, the flow can be assumed to be continuum within the insert and orifice
region.
Because we are considering only the neutral gas with an upstream temperature of 
$T_T=300$~K, we limit our study to pressure-diameters of 1--5~Torr-cm.
For the cathode geometry considered, this corresponds
to an upstream pressure, $P_T$, of 2--10~Torr (266--1330~Pa). 

We further limit our study to enclosed-keeper cathodes: the boundary between the orifice and the keeper is considered
to be a wall. All other boundary conditions remain the same as described
in Section II.

\begin{figure}[ht!]
    \centering
    \includegraphics{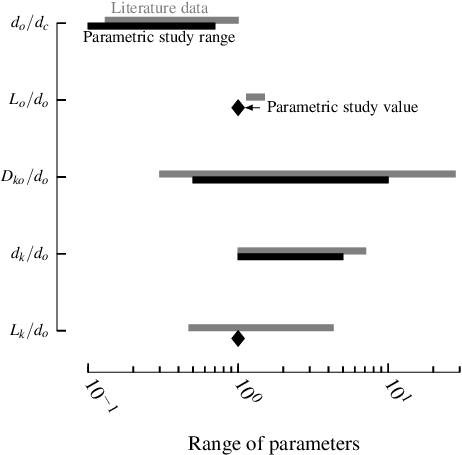}
    \caption{Range of rescaled parameters.}
    \label{fig:cathode-parameter-ranges}
\end{figure}

All dimensions may be rescaled by the inner cathode diameter, $d_c$, and the orifice diameter, $d_o$.
The range of parameters studied was determined based on the dimensions
of cathodes that are reported in the literature.
Figure~\ref{fig:cathode-parameter-ranges} 
and Table~\ref{tbl:keeper-test-cases} show the range of rescaled parameters investigated.
The cathodes for which dimensions are available, along with their respective
dimensions, are shown in Table~\ref{tbl:allcathodes-dimensions}, in the Appendix.

\begin{table}[ht!]
\caption{\label{tbl:keeper-test-cases} Range of scaled parameters considered}
\begin{ruledtabular}
\begin{tabular}{cccc}
Parameter & Unit & Symbol & Value \\
Inner channel diameter & mm & $d_c$ & 5\\
Orifice diameter & --- & $d_o/d_c$ & 0.1--0.7\\
Orifice length & --- & $L_o/d_o$ & 1\\
Keeper distance & --- & $D_{ko}/d_o$ & 0.5, 1--3, 10\\
Keeper diameter & --- & $d_k/d_o$ & 1--5\\
Keeper length & --- & $L_k/d_o$ & 1 \\
Pressure-diameter product & Torr-cm & $P_T d_c$ & 1, 5\\
Upstream pressure & Pa & $P_T$ & 266, 1330\\ 
Upstream temperature & K & $T_T$ & 300\\
Neutral species mass & amu & $M_a$ & 131.293
\end{tabular}
\end{ruledtabular}
\end{table}

The mass flow rate may be estimated by assuming a continuum
flow and a choked orifice,
\begin{equation}
    \dot{m} = C_d \dfrac{P_T \pi r_o^2}{\sqrt{T_T}}\sqrt{\dfrac{\gamma}{R_g}}
    \left(\dfrac{2}{\gamma+1}\right)^{\left(\gamma+1\right)/\left(\gamma-1\right)}
    \label{eqn:choked-mass-flow-rate}
\end{equation}
where $C_d$ is the discharge coefficient and $R_g$ the gas constant.
Using Equation~\ref{eqn:choked-mass-flow-rate} and assuming $C_d = 1$, the range of mass
flow rates considered is approximately
0.28--13.5~mg/s (2.8--135~sccm of xenon).
We limit our study to $d_o/d_c = $ 0.1 for $P_T d_c = 5$~Torr-cm: 
for this particular pressure-diameter, the corresponding mass flow
rate is above 34 mg/s (\textit{i.e.}, $\geq 300$~sccm) for $d_o/d_c = 0.5$,
which is much higher than typical cathode flow rates.

\subsection{Computational domain}

In order to decrease the overall computational cost and improve the time-to-solution 
we restrict the size of the domain, adapt the grid, and consider that 
the size of single computational cell is half that of the smallest mean free path 
in the densest regions of the simulation domain
(\textit{i.e.}, $N_\lambda = 2$).

For most cases, we do not consider the high-density insert
region and introduce particles through the orifice exit directly.
However, care must be taken in specifying the profile of the flow quantities at the orifice exit.  
For a temperature of 300~K, the Reynolds number at the orifice exit, $\textrm{Re}_o$, is on the order of $\mathcal{O}\left(10^2\right)$.
While this places the flow in the laminar regime, it is not fully
developed at the orifice exit. 
The entrance length, $L_e$, or length before which the flow is fully developed, is given
by:\cite{White}
\begin{equation}
\dfrac{L_e}{d_o} = \dfrac{0.6}{1+0.035\textrm{Re}_o} + 0.0575 \textrm{Re}_o.
\end{equation}
In all cases, $\dfrac{L_e}{d_o} > \dfrac{L_o}{d_o}$: the flow is not fully
developed.

The Mach number is typically in the range of 0.9--1.2 at the 
orifice exit\cite{Lilly2004,Varoutis2008,Titarev2014} due to slip flow, viscosity, and 2-D effects.
We have observed a Mach number at the orifice inlet, $M_{o,in}$, of $\approx 0.4$ in our simulations. 
To determine the orifice exit conditions, 
we assume that the flow 
\begin{itemize}
    \item may be modeled with a Fanno flow approach (\textit{i.e.}, the flow is in the continuum regime in the orifice), and
    \item is sonic at the orifice exit (Ma$=1$). 
\end{itemize}
The Fanno flow friction coefficient, $f$, for the orifice is given by:\cite{Anderson}
\begin{equation}
    f = \dfrac{1}{4}\dfrac{d_o}{L_o}\left( 
    \dfrac{1-M_{o,in}^2}{\gamma M_{o,in}^2}
    +\dfrac{\gamma+1}{2\gamma}
    \log\left( \dfrac{\left(\gamma+1\right) M_{o,in}^2}{2+\left(\gamma-1\right)M_{o,in}^2}\right)
    \right).
\end{equation}
For an aspect ratio $L_o/d_o =1$ and $M_{o,in}\approx 0.4$, $f\approx 0.5$.

\begin{figure*}
    \centering
    \includegraphics[scale=0.9]{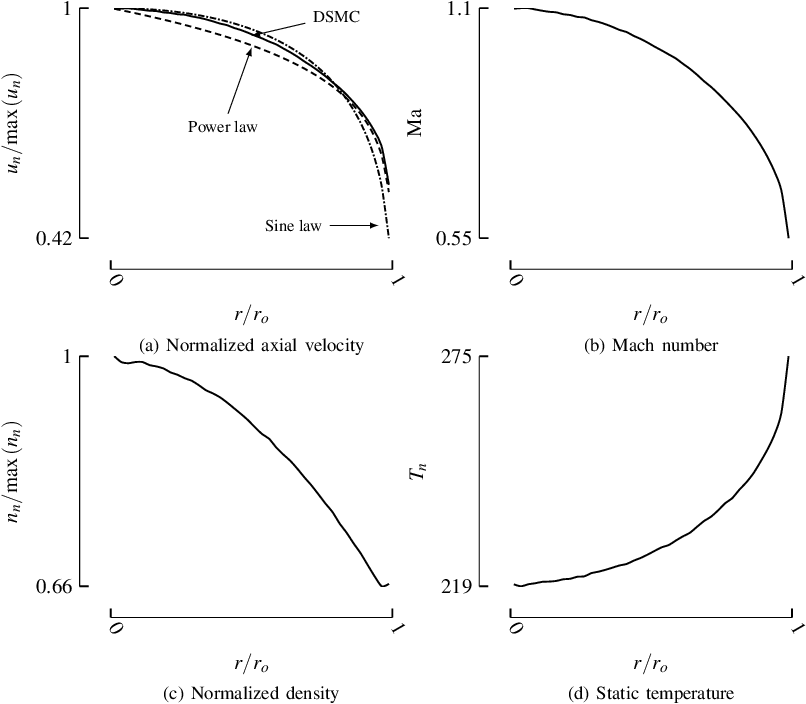}
    \caption{Normalized radial profile of (a) axial velocity, (b) Mach number, (c) number density, and (d) static temperature, for the case where $P_T d_c = 1$~Torr-cm, $d_o/d_c = 0.1$, 
$D_{ko}/d_o = 1$, and $d_k/d_o = 2$.}
    \label{fig:radial-profiles-example}
\end{figure*}

\begin{figure*}
    \centering
    \includegraphics[scale=0.8]{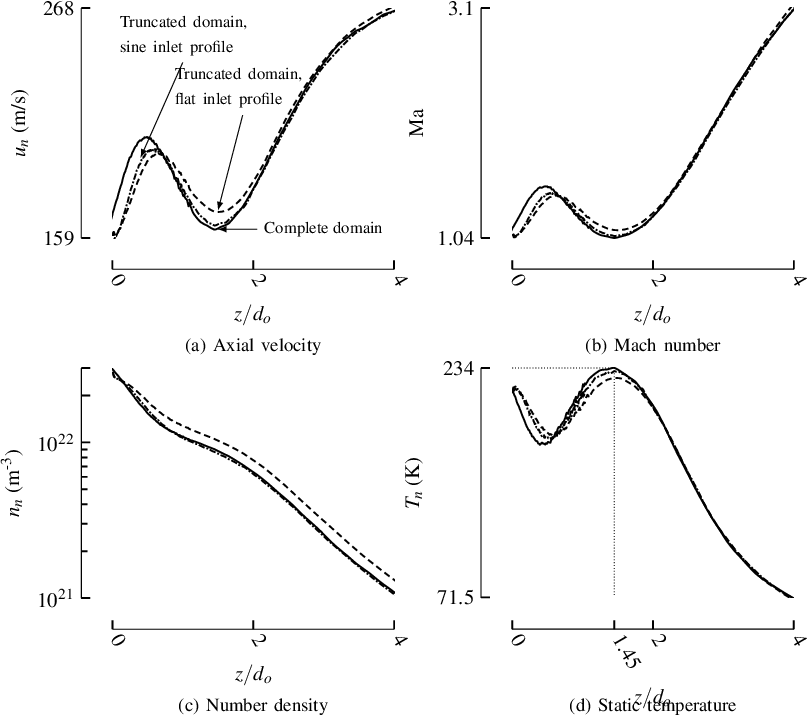}
    \caption{Center-line (a) axial velocity, (b) Mach number, (c) number density, and (d) static temperature for the case where $P_T d_c = 1$~Torr-cm, $d_o/d_c = 0.1$, $D_{ko}/d_o = 1$, and $d_k/d_o = 2$. Numerical results for both a complete domain and a truncated domain are shown. A flat-top profile for the velocity, density, and temperature is assumed for the inlet of the truncated domain.}
    \label{fig:comparison-flow-quantities-truncated-vs-full}
\end{figure*}

A ``flat-top'' velocity, temperature, and density profile is typically
not observed in orifice flows in the hydrodynamic regime.
Because the flow is not fully developed, the radial velocity profile, therefore, 
does not follow the usual square dependency on the radius that laminar flows exhibit.
Figure~\ref{fig:radial-profiles-example} shows the radial profiles of axial velocity, temperature, and density, for the case where $P_T d_c = 1$~Torr-cm, $d_o/d_c = 0.1$, 
$D_{ko}/d_o = 1$, and $d_k/d_o = 2$.
Velocity profiles typically used for turbulent flows 
fit the DSMC results reasonably well: both the ``1/7-th'' power law 
and the sinusoidal fit suggested by De Chant~\cite{DeChant2005}
show good agreement with our numerical data.
Figure~\ref{fig:comparison-flow-quantities-truncated-vs-full} shows the center-line
Mach number, pressure, density, and axial velocity, for the case where $P_T d_c = 1$~Torr-cm, $d_o/d_c = 0.1$, $D_{ko}/d_o = 1$, and $d_k/d_o = 2$.
Three cases are shown: 
\begin{itemize}
    \item a complete solution that includes the high-density insert region,
    \item a solution for a truncated domain with a flat-top profile at the cathode orifice exit (\textit{i.e.}, the inlet for this particular case), and 
    \item a solution for a truncated domain with an imposed, non-flat velocity and temperature profile at the cathode orifice exit.
\end{itemize}
The numerical results for the simplified domain are qualitatively similar to that of the complete domain.
However, the Mach number, pressure, and velocity are all under-estimated near the
cathode orifice exit ($x/d_o < 1$) for the truncated domain cases.
This is likely a result of \textit{(i)} the radial velocity being neglected near the orifice wall
and \textit{(ii)} the orifice exit density being underestimated with the Fanno flow approach.
However, good agreement is obtained with the full solution for $x/d_o > 0.5$ when 
the inlet velocity follows a sinusoidal radial profile.

For the truncated domain cases, particles will, therefore, be introduced in the keeper region directly from the orifice. 
The velocity profile will be assumed to be sinusoidal.
Assuming a friction coefficient, $f$, equal to 0.5, and an orifice 
exit Mach number of one,
the inlet flow quantities are determined from the upstream stagnation quantities 
through the following procedure:
\begin{enumerate}
    \item Using a Fanno flow approach, solve for Mach number
    and relevant pressure and temperature ratios at the orifice entrance. 
    \item Compute the pressure, temperature at the orifice exit 
    from the ratios obtained in Step 1 and the assumed Mach number of 1 at the orifice exit
    \item Compute the radial velocity profile using the approach suggested by 
    DeChant\cite{DeChant2005}
    \item Compute the radial static temperature profile from the velocity profile and
    the definition of the Mach number.
\end{enumerate}

\section{Results and discussion}

\subsection{Flow structure in the presence of a keeper electrode}

Accurately resolving 
the neutral flow field is critical for comprehensive plasma-neutral cathode simulations 
that use the DSMC method for the neutrals.
The introduction of an enclosed keeper strongly affects the
neutral flow downstream of the orifice.  
The flow structure differs from that of a simple expansion from 
the cathode orifice into vacuum (see, \textit{e.g.}, Figure~\ref{fig:cao-contours}(a)). 
In the presence of the keeper, the flow through the orifice first 
expands into a larger channel
(\textit{i.e.}, the orifice-keeper region). After reaching the keeper,
the flow undergoes a second expansion, into vacuum. 
Our numerical results reveal the existence of two main flow regimes: the flow
may be either supersonic or subsonic in the orifice-keeper region,
depending on the value of keeper-to-total pressure ratio, $P_{ko}/P_T$.
The ``keeper pressure'', $P_{ko}$, is defined as the average (static) pressure in the 
region bounded by the orifice plate and keeper entrance.
Only the quiescent part of the flow (\textit{i.e.}, $v\approx 0$ m/s) is considered
for the computation of $P_{ko}$.
$P_T$ is the total upstream pressure.
The pressure ratio we define is similar to the ``ambient-to-total'' (or ``reservoir-to-total'') pressure ratio
that is typically used to characterize nozzle flows.

\subsubsection{Supersonic flow}
\begin{figure*}
    \centering
\scalebox{0.85}{
    \begin{tikzpicture}
    \node at (0,0) {\includegraphics[width=6.8in]{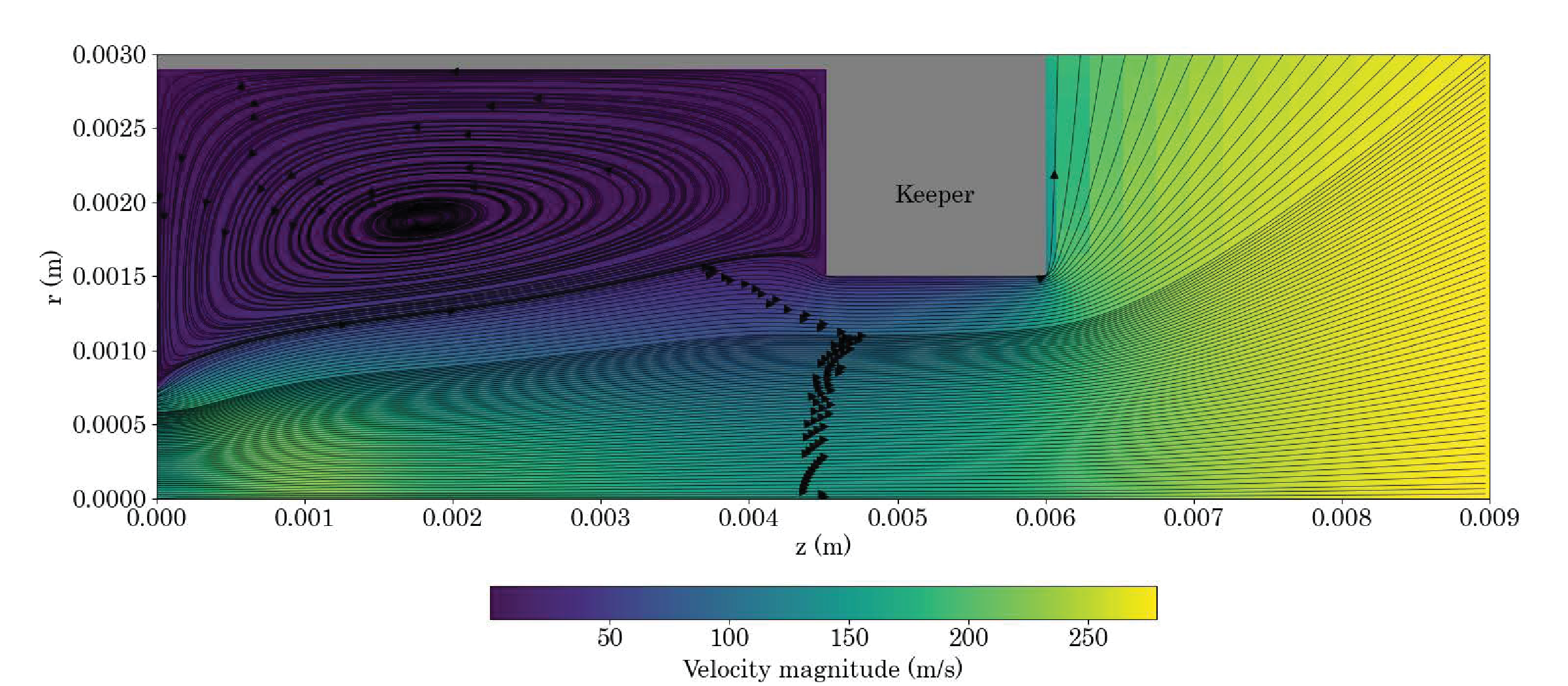}};
    \draw[latex-latex,thick] (-8.7,-1.7) -- (-8.7,-0.48);
    \node at (-8.3,-1) {\footnotesize $d_o/2$};
    \draw[latex-latex,thick] (-11.8,-1.7) -- (-11.8,2.35);
    \node at (-11.3,0.35) {\footnotesize $d_c/2$};
    
    \node at (-8.8,0.975) {\footnotesize Orifice};
    \node at (-8.8,0.675) {\footnotesize plate};
    
    \draw [thick, draw=black, fill=gray, opacity=0.2]
    (-9.4,-0.48) -- (-6.9,-0.48) -- (-6.9,3.16) -- (-12.0,3.16) -- (-12.0,2.35) -- (-9.4,2.35) -- cycle;
    \node at (-11.0,2.75) {\footnotesize Cathode tube};
    \end{tikzpicture}
}
    \caption{Velocity magnitude with streamlines for the case $P_T d_c = 1$~Torr-cm, $d_o/d_c = 0.3$, $D_{ko}/d_o = 3.0$, and $d_k/d_o = 2.0$.}
    \label{fig:typical-flow-structure-recirculation}
\end{figure*}

\begin{figure*}
    \centering
\scalebox{0.85}{
    \begin{tikzpicture}
    \node at (0,0) {\includegraphics[width=6.8in]{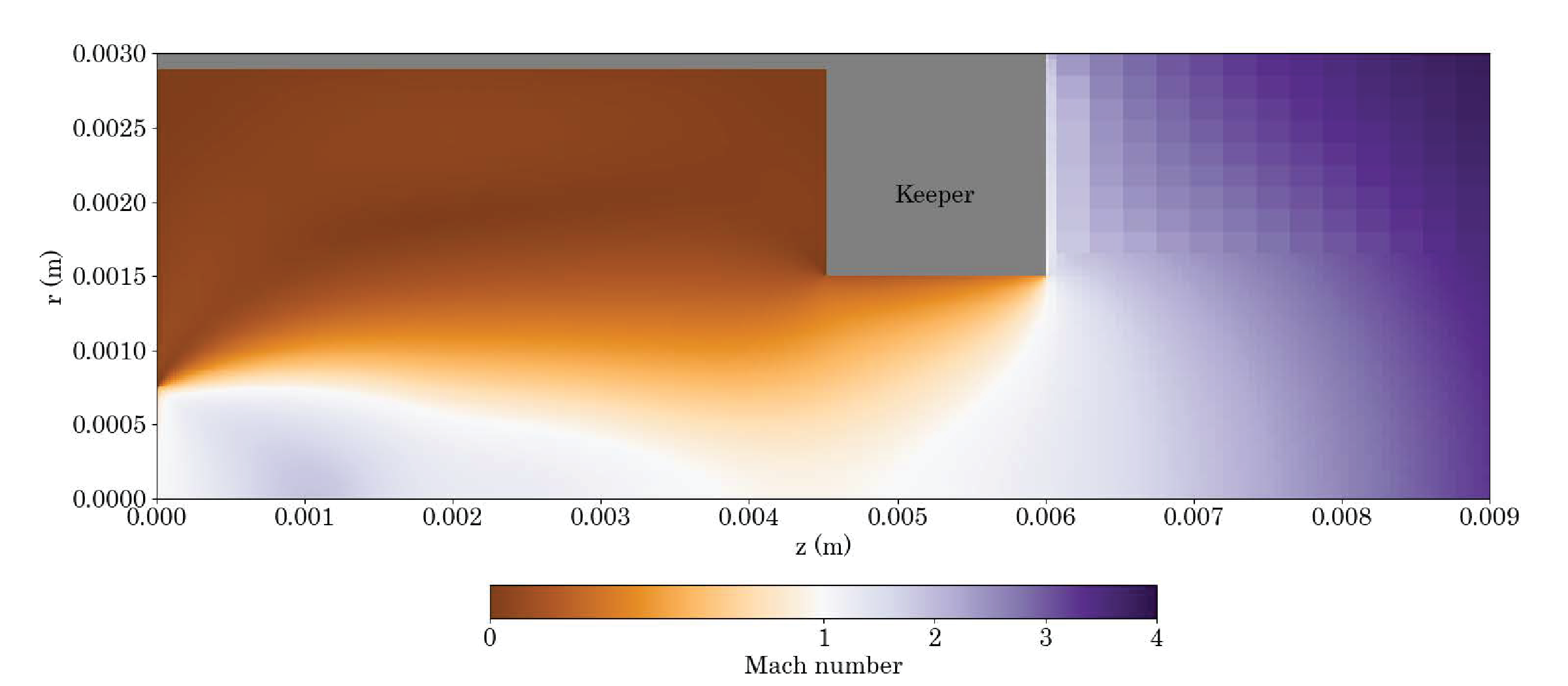}};
    \draw[latex-latex,thick] (-8.7,-1.7) -- (-8.7,-0.48);
    \node at (-8.3,-1) {\footnotesize $d_o/2$};
    \draw[latex-latex,thick] (-11.8,-1.7) -- (-11.8,2.35);
    \node at (-11.3,0.35) {\footnotesize $d_c/2$};
    
    \node at (-8.8,0.975) {\footnotesize Orifice};
    \node at (-8.8,0.675) {\footnotesize plate};
    
    \draw [thick, draw=black, fill=gray, opacity=0.2]
    (-9.4,-0.48) -- (-6.9,-0.48) -- (-6.9,3.16) -- (-12.0,3.16) -- (-12.0,2.35) -- (-9.4,2.35) -- cycle;
    \node at (-11.0,2.75) {\footnotesize Cathode tube};
    \end{tikzpicture}
}
    \caption{Mach number for the case $P_T d_c = 1$~Torr-cm, $d_o/d_c = 0.3$, $D_{ko}/d_o = 3.0$, and $d_k/d_o = 2.0$.}
    \label{fig:typical-flow-structure-Mach}
\end{figure*}

\begin{figure*}
    \centering
\scalebox{0.85}{
    \begin{tikzpicture}
    \node at (0,0) {\includegraphics[width=6.8in]{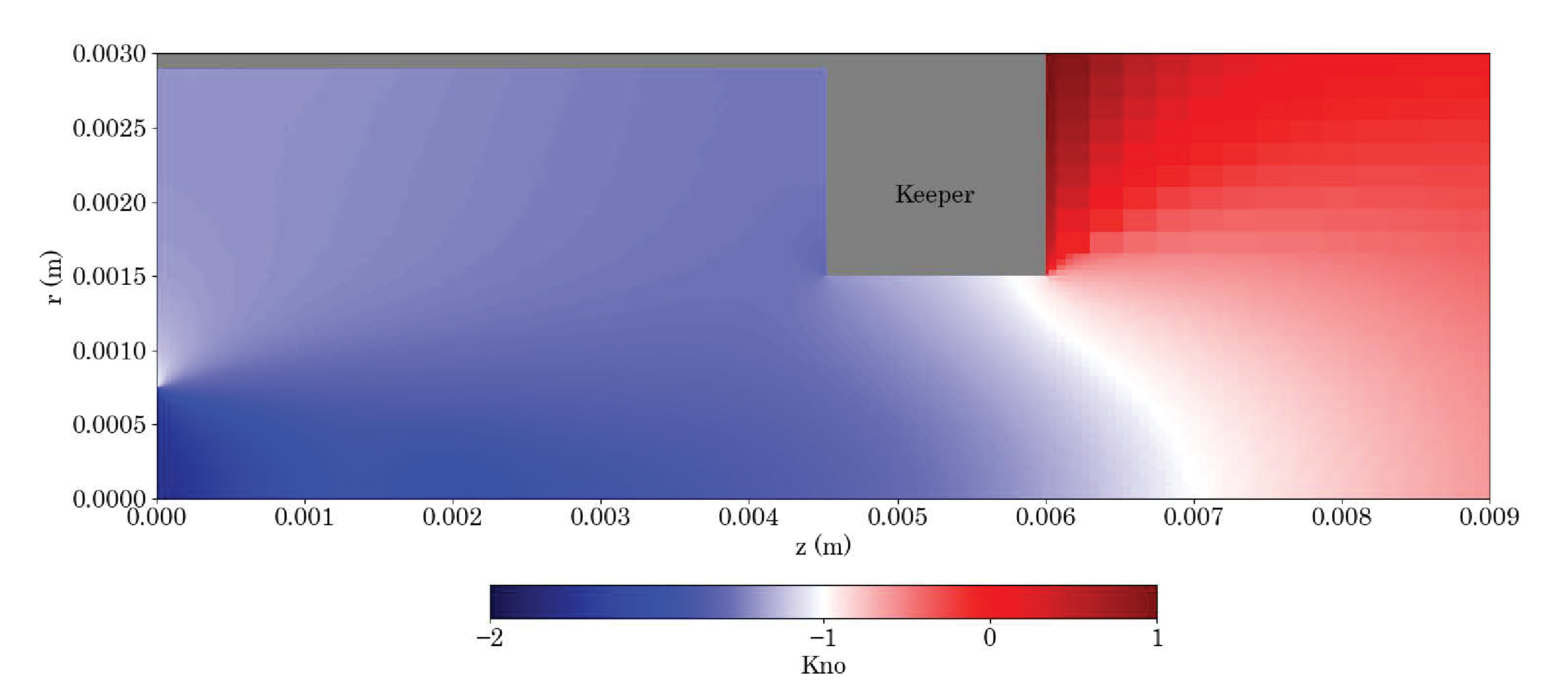}};
    \draw[latex-latex,thick] (-8.7,-1.7) -- (-8.7,-0.48);
    \node at (-8.3,-1) {\footnotesize $d_o/2$};
    \draw[latex-latex,thick] (-11.8,-1.7) -- (-11.8,2.35);
    \node at (-11.3,0.35) {\footnotesize $d_c/2$};
    
    \node at (-8.8,0.975) {\footnotesize Orifice};
    \node at (-8.8,0.675) {\footnotesize plate};
    
    \draw [thick, draw=black, fill=gray, opacity=0.2]
    (-9.4,-0.48) -- (-6.9,-0.48) -- (-6.9,3.16) -- (-12.0,3.16) -- (-12.0,2.35) -- (-9.4,2.35) -- cycle;
    \node at (-11.0,2.75) {\footnotesize Cathode tube};

    \node[fill=white] at (0.2,-3.6) {$\log_{10}\left(\textrm{Kn}_o\right)$};
    \end{tikzpicture}
}
    \caption{Knudsen number normalized by orifice diameter for the case $P_T d_c = 1$~Torr-cm, $d_o/d_c = 0.3$, $D_{ko}/d_o = 3.0$, and $d_k/d_o = 2.0$.}
    \label{fig:typical-flow-structure-knudsen}
\end{figure*}

\begin{figure*}
    \centering
\scalebox{0.85}{
    \begin{tikzpicture}
    \node at (0,0) {\includegraphics[width=6.8in]{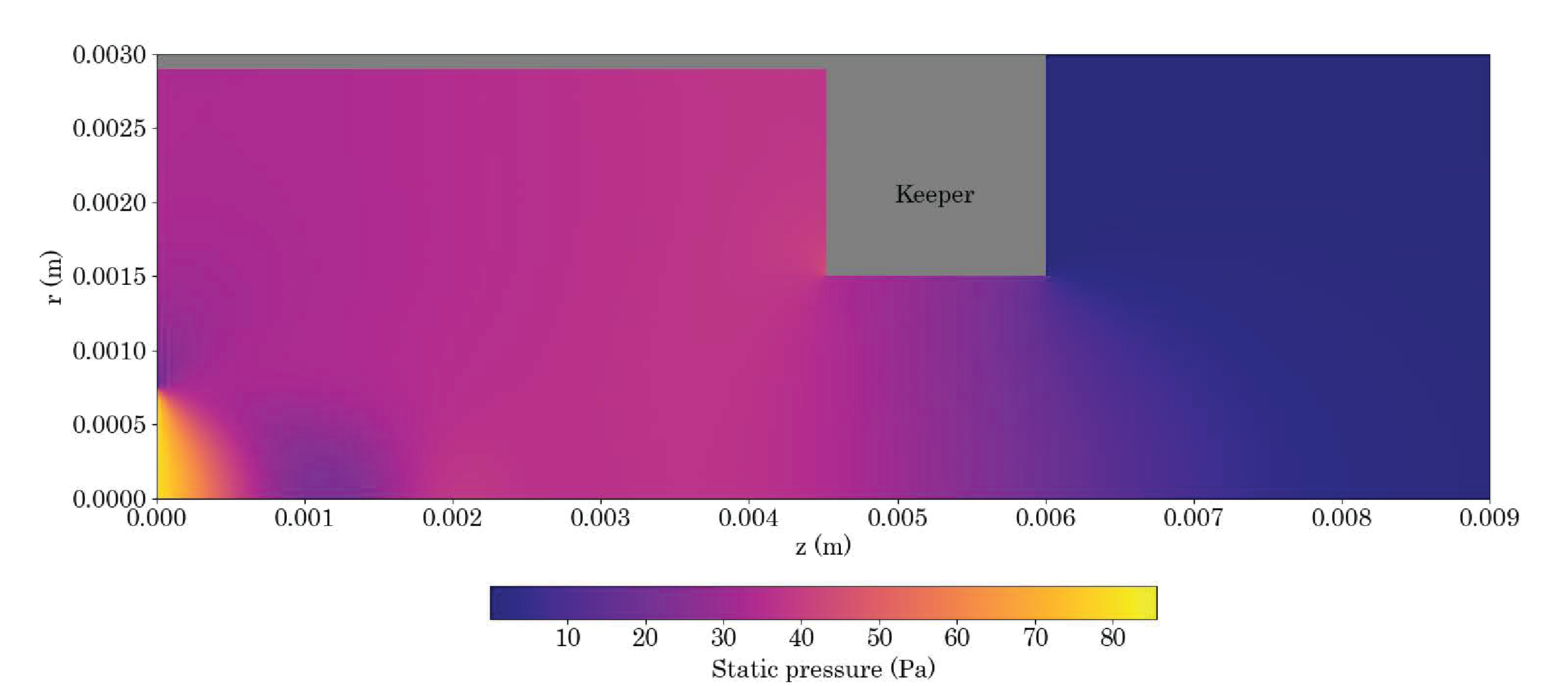}};
    \draw[latex-latex,thick] (-8.7,-1.7) -- (-8.7,-0.48);
    \node at (-8.3,-1) {\footnotesize $d_o/2$};
    \draw[latex-latex,thick] (-11.8,-1.7) -- (-11.8,2.35);
    \node at (-11.3,0.35) {\footnotesize $d_c/2$};
    
    \node at (-8.8,0.975) {\footnotesize Orifice};
    \node at (-8.8,0.675) {\footnotesize plate};
    
    \draw [thick, draw=black, fill=gray, opacity=0.2]
    (-9.4,-0.48) -- (-6.9,-0.48) -- (-6.9,3.16) -- (-12.0,3.16) -- (-12.0,2.35) -- (-9.4,2.35) -- cycle;
    \node at (-11.0,2.75) {\footnotesize Cathode tube};
    
    \end{tikzpicture}
}
    \caption{Static pressure for the case $P_T d_c = 1$~Torr-cm, $d_o/d_c = 0.3$, $D_{ko}/d_o = 3.0$, and $d_k/d_o = 2.0$.}
    \label{fig:typical-flow-structure-pressure}
\end{figure*}

In most cases, the pressure ratio $P_{ko}/P_T$ is such that the flow that leaves 
the cathode orifice is choked and supersonic, and the
overall flow structure is similar to that reported by Titarev and Shakov\cite{Titarev2014} 
for a composite pipe system composed of two pipes of increasing radii that exhaust
into vacuum. 
Figures~\ref{fig:typical-flow-structure-recirculation} through~\ref{fig:typical-flow-structure-pressure} illustrate the typical flow structure in
the case where the orifice is choked.
The flow expansion from the cathode orifice to 
the keeper-orifice region results in an underexpanded, rarefied, supersonic jet
(Figures~\ref{fig:typical-flow-structure-recirculation},~\ref{fig:typical-flow-structure-Mach},~\ref{fig:typical-flow-structure-pressure}).
After exiting the orifice, the jet interacts with the keeper plate and a 
toroidal zone of recirculation
appears in the region between the underexpanded jet and the top boundary 
(Figure~\ref{fig:typical-flow-structure-recirculation}).
The Knudsen number in the keeper region is $\approx 0.1$ (Figure~\ref{fig:typical-flow-structure-knudsen}), indicating that the flow
is in the transitional regime. 

\begin{figure*}
    \centering
    \includegraphics[width=6.8in]{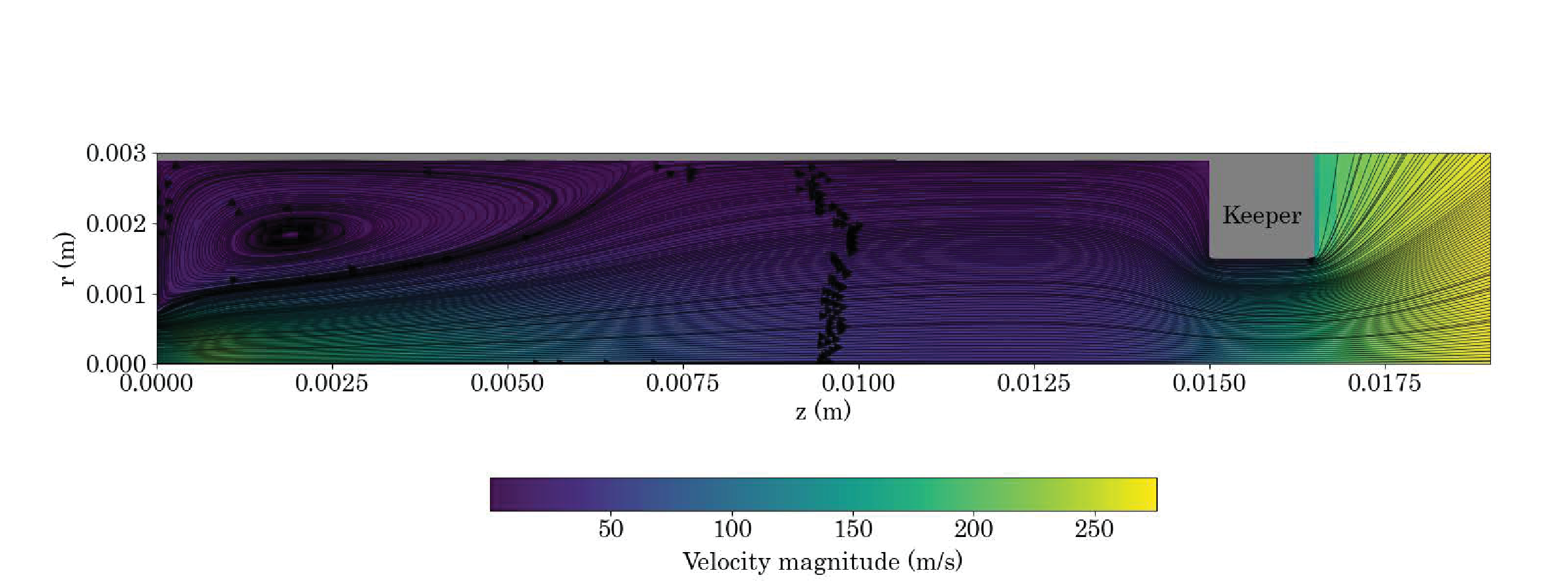}
    \caption{Velocity magnitude with streamlines for the case $P_T d_c = 1$~Torr-cm, $d_o/d_c = 0.3$, $D_{ko}/d_o = 10.0$, and $d_k/d_o = 2.0$.}
    \label{fig:typical-flow-structure-Dkodo-10-recirculation}
\end{figure*}

\begin{figure*}
    \centering
\includegraphics[scale=0.97]{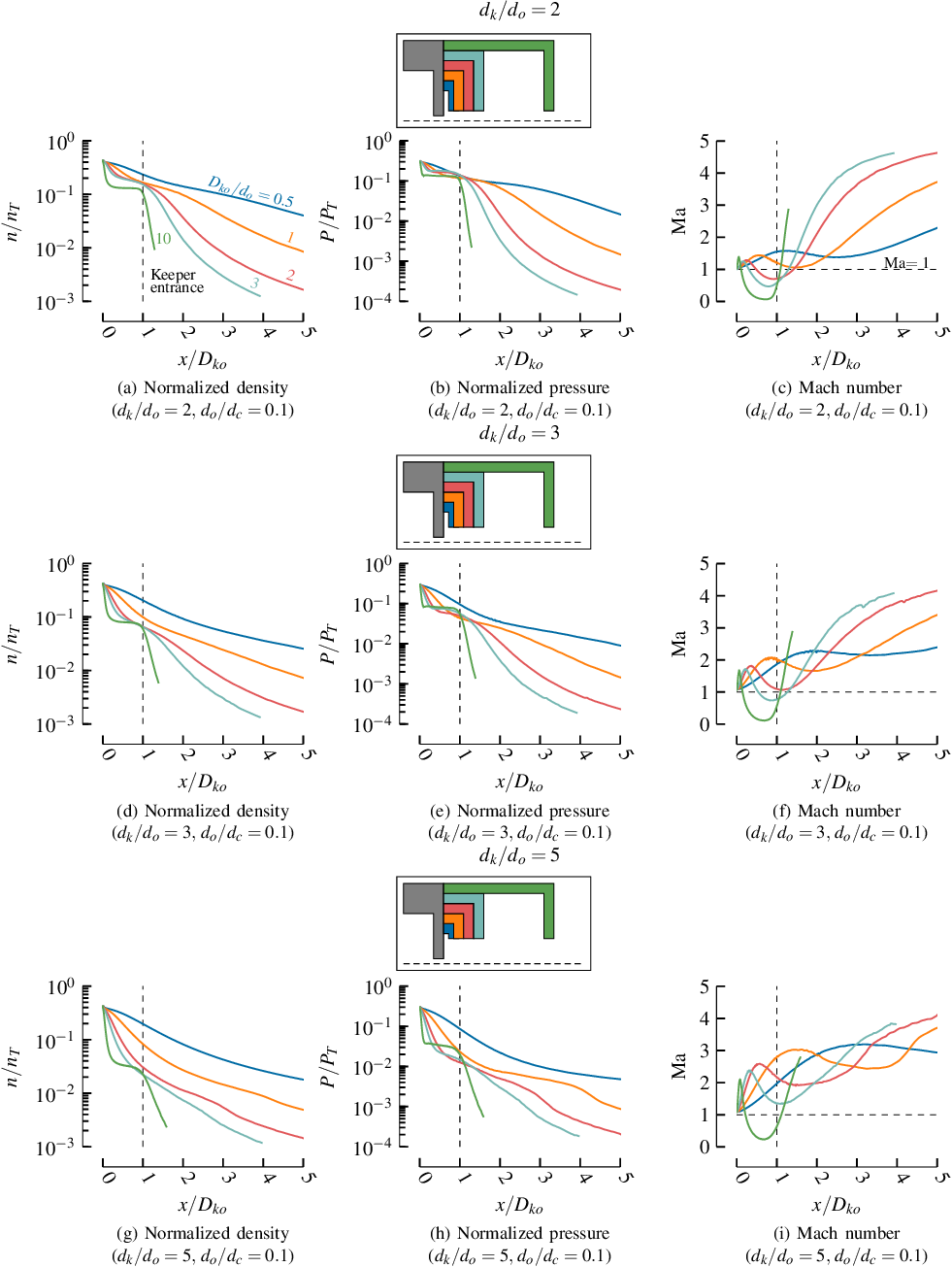}
    \caption{Centerline values for $d_o/d_c = 0.1$ and $P_T d_c = 1$~Torr-cm.
    (a,d,g) Normalized density, (b,e,h) normalized pressure, (c,f,i) Mach number. 
    Density and pressure
    are normalized by the upstream values ($n_T=6.4\cdot 10^{22}$~m\textsuperscript{-3} and $P_T = 267$~Pa, respectively). 
    The horizontal axis is normalized by the orifice-keeper distance, $D_{ko}$.
    {A scaled schematic of each configuration investigated is shown under each $d_k/d_o$ case. The keeper entrance for $D_{ko}/d_o = 1$ overlaps the keeper exit for $D_{ko}/d_o = 0.5$}}
    \label{fig:centerline-dodc-0.1}
\end{figure*}

\begin{figure*}
    \centering
\includegraphics[scale=0.97]{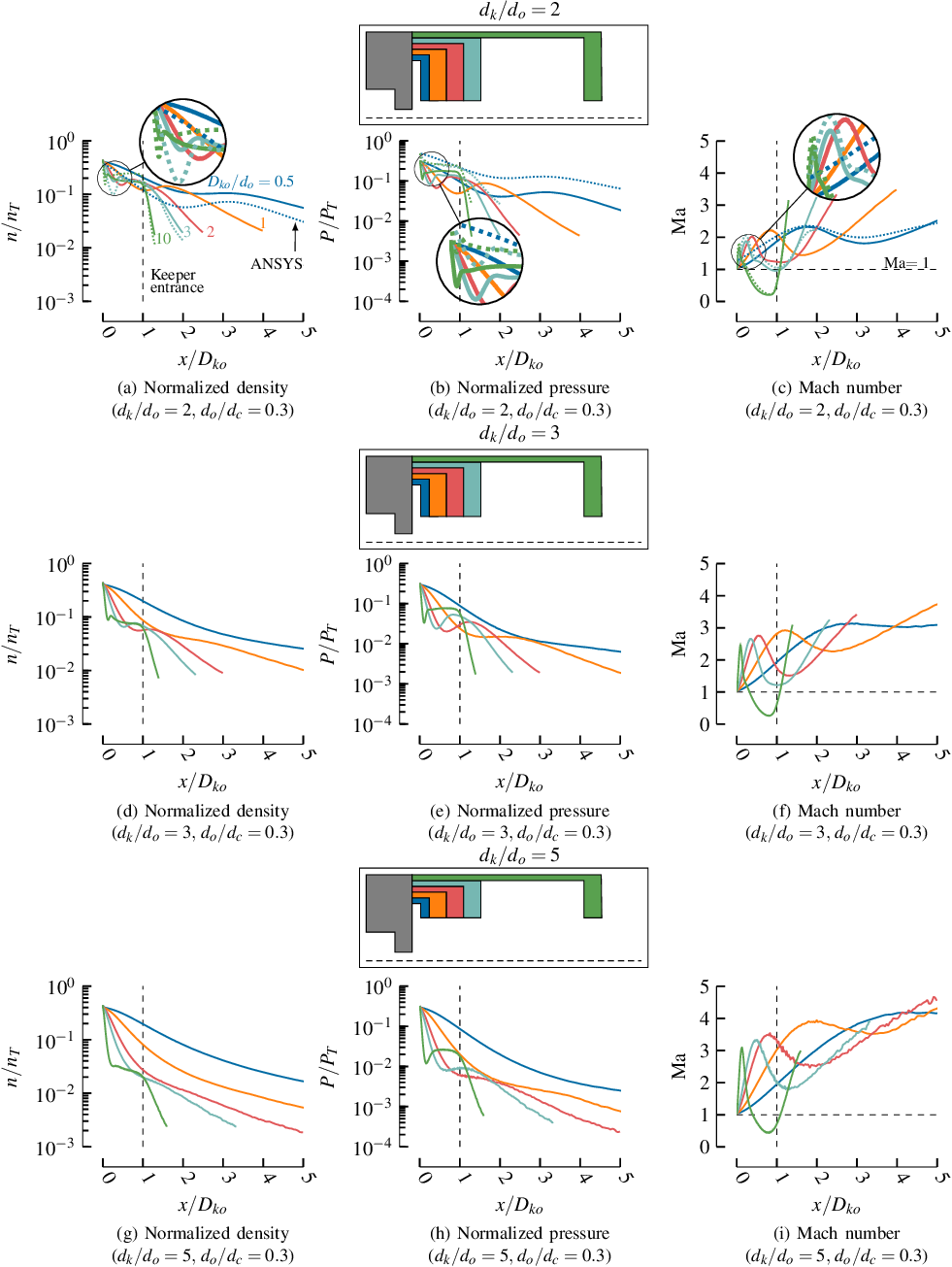}
    \caption{Centerline values for $d_o/d_c = 0.3$ and $P_T d_c = 1$~Torr-cm.
    (a,d,g) Normalized density, (b,e,h) normalized pressure, (c,f,i) Mach number.
    Density and pressure
    are normalized by the upstream values ($n_T=6.4\cdot 10^{22}$~m\textsuperscript{-3} and $P_T = 267$~Pa, respectively). 
    The horizontal axis is normalized by the orifice-keeper distance, $D_{ko}$.
    {A scaled schematic of each configuration investigated is shown under each $d_k/d_o$ case. The keeper entrance for $D_{ko}/d_o = 1$ overlaps the keeper exit for $D_{ko}/d_o = 0.5$}}
    \label{fig:centerline-dodc-0.3}
\end{figure*}

\begin{figure*}
    \centering
\includegraphics[scale=0.97]{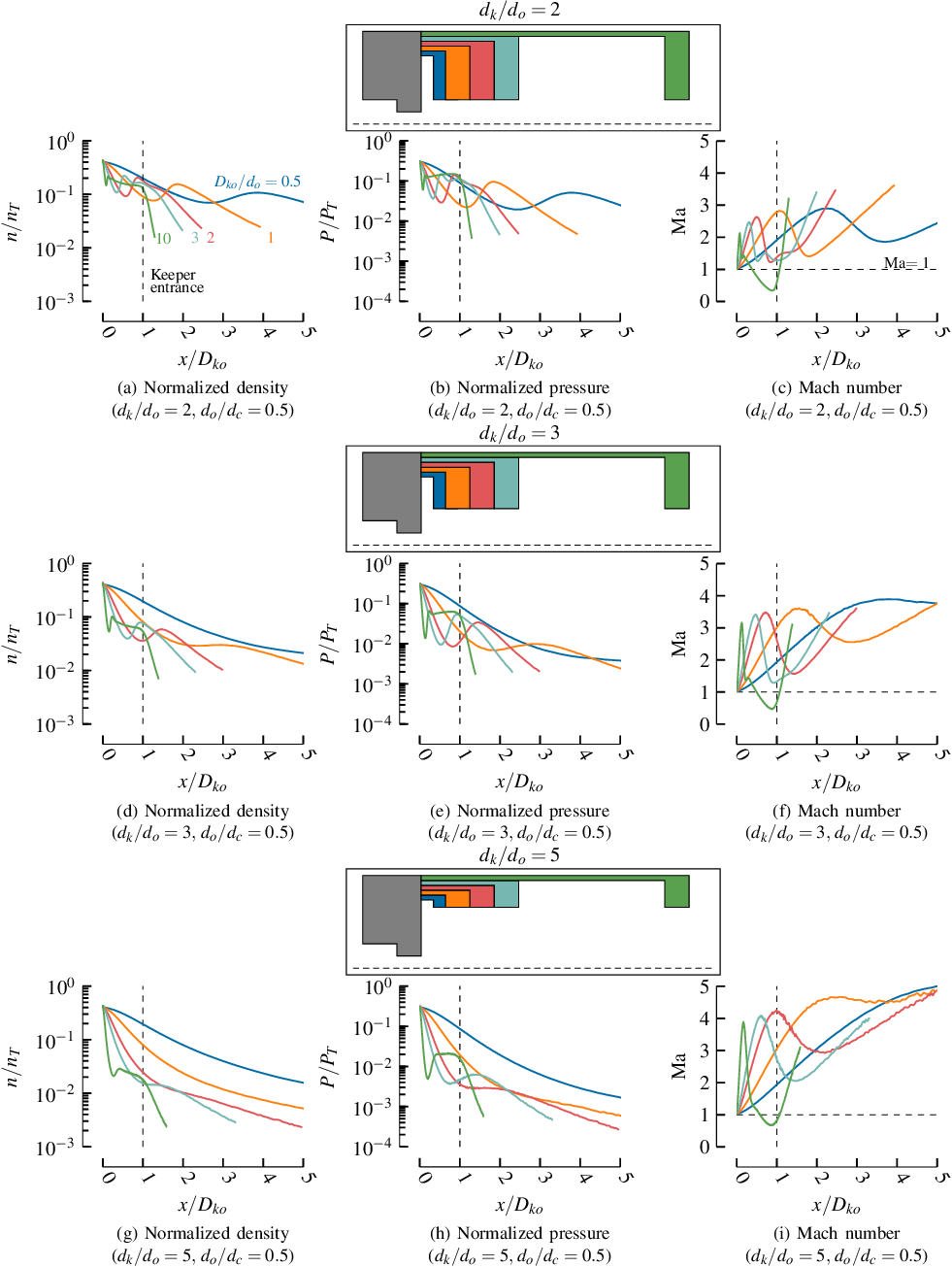}
    \caption{Centerline values for $d_o/d_c = 0.5$ and $P_T d_c = 1$~Torr-cm.
    (a,d,g) Normalized density, (b,e,h) normalized pressure, (c,f,i) Mach number.
    Density and pressure
    are normalized by the upstream values ($n_T=6.4\cdot 10^{22}$~m\textsuperscript{-3} and $P_T = 267$~Pa, respectively). 
    The horizontal axis is normalized by the orifice-keeper distance, $D_{ko}$.
    {A scaled schematic of each configuration investigated is shown under each $d_k/d_o$ case. The keeper entrance for $D_{ko}/d_o = 1$ overlaps the keeper exit for $D_{ko}/d_o = 0.5$}}
    \label{fig:centerline-dodc-0.5}
\end{figure*}

\begin{figure*}
    \centering
\includegraphics[scale=0.97]{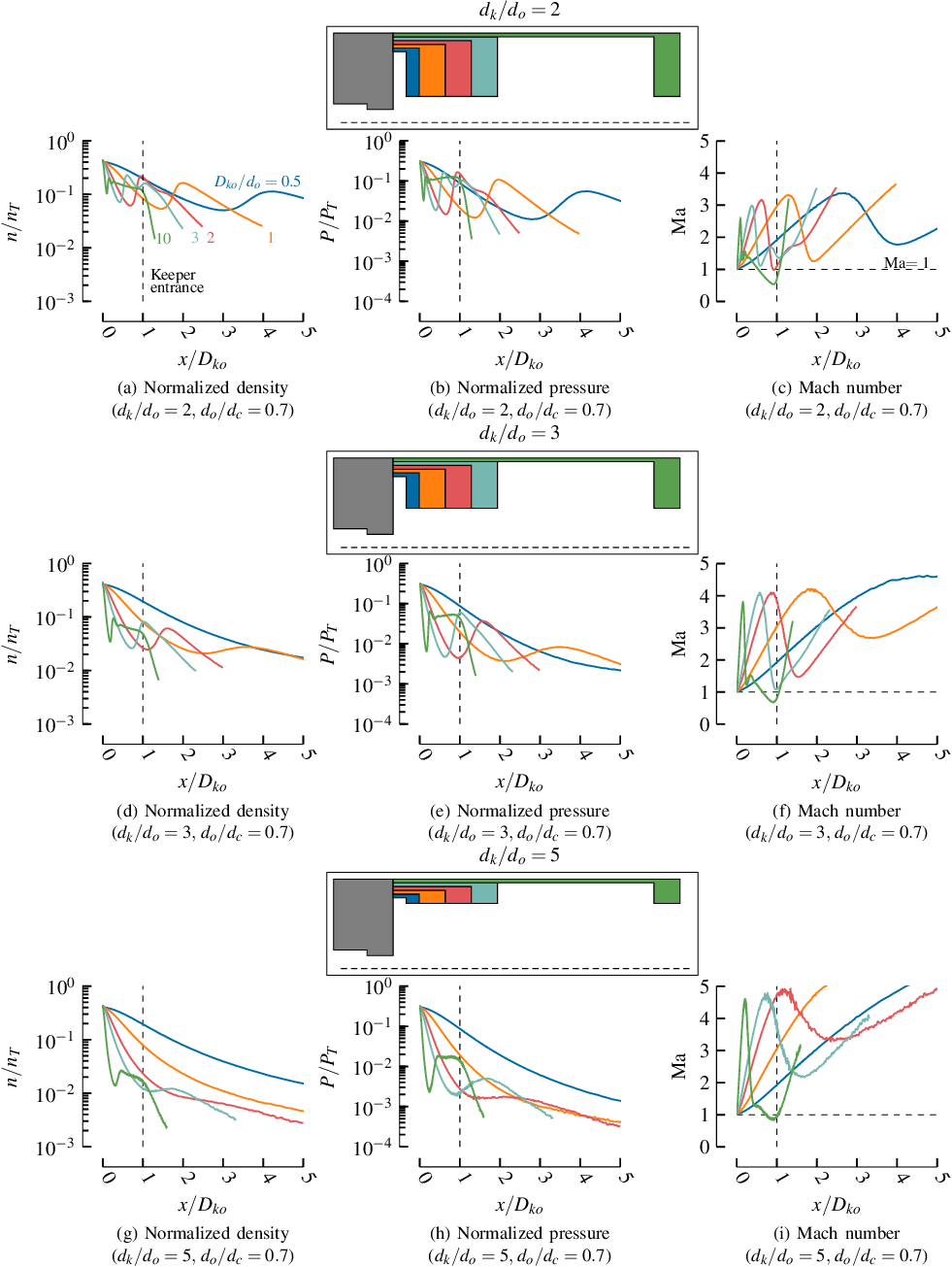}
    \caption{Centerline values for $d_o/d_c = 0.7$ and $P_T d_c = 1$~Torr-cm.
    (a,d,g) Normalized density, (b,e,h) normalized pressure, (c,f,i) Mach number.
    Density and pressure
    are normalized by the upstream values ($n_T=6.4\cdot 10^{22}$~m\textsuperscript{-3} and $P_T = 267$~Pa, respectively). 
    The horizontal axis is normalized by the orifice-keeper distance, $D_{ko}$.
    {A scaled schematic of each configuration investigated is shown under each $d_k/d_o$ case. The keeper entrance for $D_{ko}/d_o = 1$ overlaps the keeper exit for $D_{ko}/d_o = 0.5$}}
    \label{fig:centerline-dodc-0.7}
\end{figure*}

The exact structure of the underexpanded jet depends on the 
backing-to-upstream pressure ratio, and, therefore, the cathode geometry.
The centerline Mach number, static pressure, and number density are shown
in Figures~\ref{fig:centerline-dodc-0.1} through~\ref{fig:centerline-dodc-0.7}
for $d_o/d_c = $ 0.1--0.7, respectively.
We observe the following qualitative trends:
\begin{itemize}
    \item For short orifice-to-keeper distances and larger keeper-to-orifice ratios ($D_{ko} /d_o \leq 1$, $d_k / d_o \geq 3$), the orifice and keeper expansions 
    merge to form a single expansion system (see, \textit{e.g.}, Figure~\ref{fig:centerline-dodc-0.3}(d) for $D_{ko} / d_o = 0.5$).
    \item For the longest orifice-to-keeper distances ($D_{ko}/d_o = 10$), 
    the orifice expansion results in a subsonic flow in the orifice-keeper region,
    and the recirculation zone remains confined near the cathode orifice plate (Figure~\ref{fig:typical-flow-structure-Dkodo-10-recirculation}).
    The flow transitions from supersonic to subsonic in the orifice-keeper region
    and a plateau in both density and pressure exist. The transition may be through
    an expansion ($d_o/d_c = 0.1$) or through a diffuse shock ($d_o/d_c \geq 0.3$).
    In the latter case, the static pressure increases past the shock (see, \textit{e.g.}, Figure~\ref{fig:centerline-dodc-0.5}(a--c) for $x/D_{ko} \approx 0.3$).
    \item As the keeper-to-orifice diameter ratio, $d_k/d_o$, increases, the keeper expansion
    into vacuum becomes more pronounced: at the keeper entrance, the Mach number
    increases with $d_k/d_o$, 
    while both static pressure and number density decrease
    with $d_k/d_o$ (see, \textit{e.g.}, Figures~\ref{fig:centerline-dodc-0.7}(a,d,g))
    \item  For large enough mass flow rates (\textit{i.e.}, 
    large enough orifice-to-cathode diameter ratio since $P_T$ is fixed), multiple
    cells of the underexpanded jet appear and a Mach disk may form (see, \textit{e.g.}, Figure~\ref{fig:centerline-dodc-0.7}(c), for $D_{ko}/d_o = 3$).
\end{itemize}
Figure~\ref{fig:centerline-dodc-0.3}(a--c) show a comparison of the DSMC numerical results
to that produced by ANSYS-CFX, for select cases ($d_o/d_c = 0.3$, $d_k/d_o = 2$, and $D_{ko}/d_o =0.5,\,3,\,\textrm{and}\,10$). 
Results from the continuum flow solver qualitatively reproduce that of the DSMC solver.
However, the values for the density (and pressure) 
are consistently underestimated (resp., overestimated), and both density and 
pressure decay rates in the keeper and plume regions are inconsistent with that predicted 
by the DSMC approach.
The continuum flow solver solution was obtained on a computational domain that extends upstream in 
the insert region, which was not covered by the DSMC analysis. 
Therefore, the inlet parameters of the DSMC analysis presented on
Figure~\ref{fig:radial-profiles-example} differ from the profiles obtained with a 
consistent analysis that uses a continuum model. 
Better closer quantitative agreement between the DSMC and continuum models may be obtained 
if the DSMC analysis is performed using initial profiles at the nozzle that correspond to the 
results of the continuum flow analysis.
This approach is beyond the scope of this work.

\begin{figure}[ht!]
    \centering
    \includegraphics{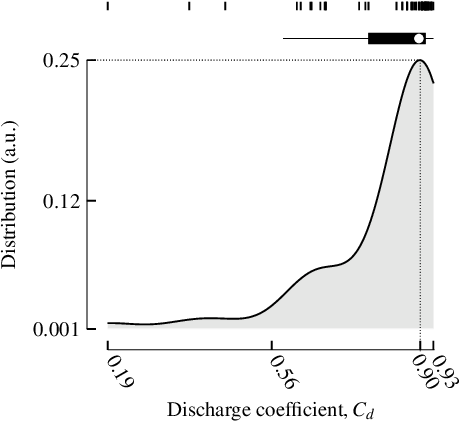}
    \caption{Combined violin and box plot of the range of discharge coefficient, $C_d$.
    Each individual tick mark indicates the location of a data point.}
    \label{fig:discharge-coefficient-distribution}
\end{figure}

A statistical distribution of the discharge coefficient as obtained with the method outlined in Section II.C.1.c
is shown in Figure~\ref{fig:discharge-coefficient-distribution} for the supersonic flow cases.
The mass flow rate as obtained for the isentropic mass flow rate, $\dot{m}_{th}$, that is
required to compute $C_d$ is given by
\begin{equation}
    \dot{m}_{th} = \dfrac{P_T \pi r_o^2}{\sqrt{T_T}}\sqrt{\dfrac{\gamma}{R_g}}
    \left(\dfrac{2}{\gamma+1}\right)^{\left(\gamma+1\right)/\left(\gamma-1\right)}
    .
\end{equation}
The average value and 95\% confidence bounds for $C_d$ are 0.82, and 0.55 and 0.92, respectively.
The study of the variation of $C_d$ with the geometric parameters (\textit{i.e.}, $d_o/d_c$, $d_k/d_o$, and $D_{ko}/d_o$) is beyond the scope of this article.
For a cathode with a similar geometry as the one studied and for a set upstream
pressure, a range of mass flow rate may be estimated with 
Equation~\ref{eqn:choked-mass-flow-rate} and $C_d$ in the 95\% confidence interval.
This approach may also be used to estimate the upstream (total) pressure of the gas, $P_T$, for a given
mass flow rate.

\subsubsection{Subsonic flow}

\begin{figure*}
    \centering
    \scalebox{0.8}{
    \begin{tikzpicture}
    \node at (0,0) {\includegraphics[width=6.8in]{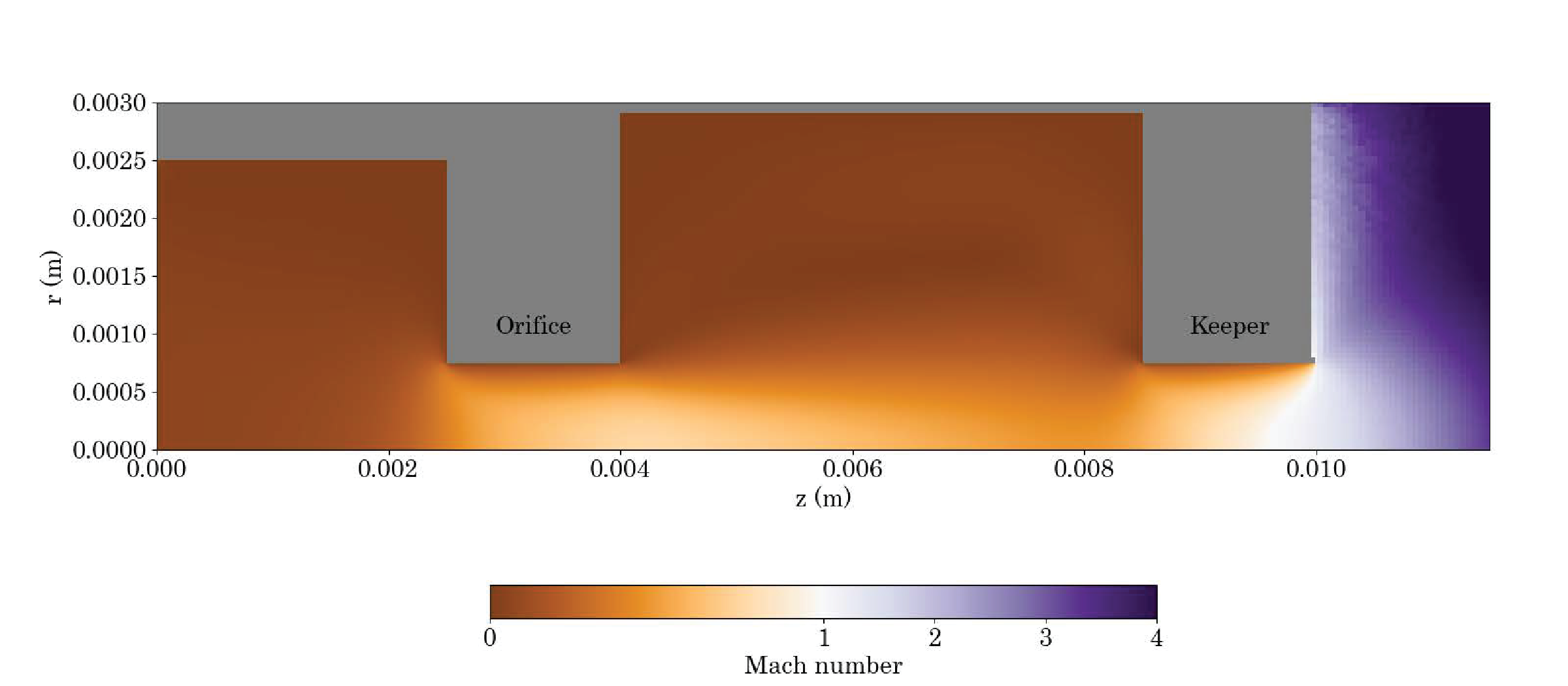}};
    \end{tikzpicture}
    }
    \caption{Mach number for the (subsonic) case $P_T d_c = 1$~Torr-cm, $d_o/d_c = 0.3$, $D_{ko}/d_o = 3.0$, and $d_k/d_o = 1.0$.}
    \label{fig:typical-flow-structure-Mach-subsonic}
\end{figure*}

\begin{figure*}
    \centering
    \scalebox{0.8}{
    \begin{tikzpicture}
    \node at (0,0) {\includegraphics[width=6.8in]{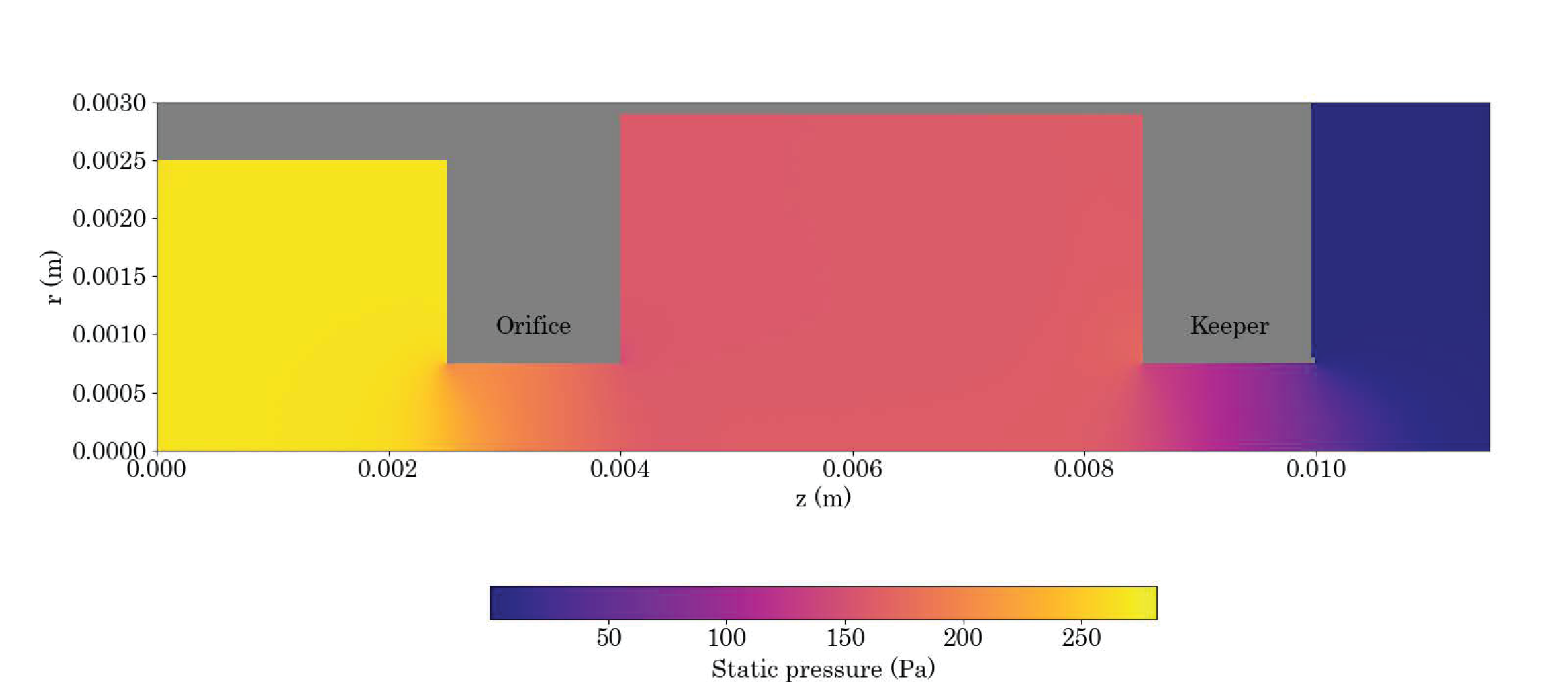}};
    \end{tikzpicture}
    }
    \caption{Static pressure for the (subsonic) case $P_T d_c = 1$~Torr-cm, $d_o/d_c = 0.3$, $D_{ko}/d_o = 3.0$, and $d_k/d_o = 1.0$.}
    \label{fig:typical-flow-structure-pressure-subsonic}
\end{figure*}

In some cases where $d_o = d_k$, the keeper induces a 
(static) pressure rise that is large enough to prevent the orifice
from choking.
In this configuration the flow remains subsonic throughout the orifice-keeper region,
and the supersonic expansion only occurs in the \textit{keeper} orifice (as opposed to the
main cathode orifice).
Because the flow is subsonic in this particular case, the keeper-to-total pressure ratio, $P_{ko}/P_T$, is higher than that in the supersonic case.
Figures~\ref{fig:typical-flow-structure-Mach-subsonic} and~\ref{fig:typical-flow-structure-pressure-subsonic} show
a contour of the Mach number and static pressure for the subsonic case, respectively. 
For high-enough mass flow rates, however, the flow may be supersonic. 
The exact point at which the flow transitions from subsonic to supersonic
for the case where $d_o = d_k$ is beyond the scope of this article.

\begin{figure*}
    \centering
    \scalebox{0.8}{
    \begin{tikzpicture}
    \node at (0,0) {\includegraphics{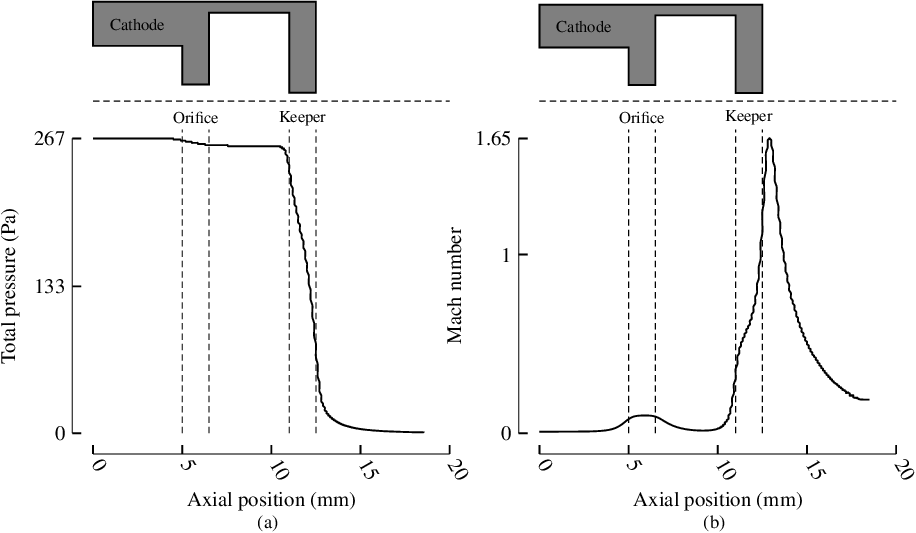}};
    \end{tikzpicture}
    } 
    \caption{(a) Total pressure and (b) Mach number for the case $d_o/d_c = 0.3$, $D_{ko}/d_o = 3$, and $d_k/d_o=0.5$.}
    \label{fig:example-dkdo-0.5}
\end{figure*}

{
Some heaterless cathodes may have a configuration for which $d_k < d_o$ (\textit{i.e.}, $d_k/d_o < 1$).
This configuration is beyond the scope of this article: a continuum or near-continuum solver 
is likely more appropriate for this case as the 
flow will remain subsonic and in the continuum regime throughout the orifice-keeper region.
This case is illustrated in Figure~\ref{fig:example-dkdo-0.5}, which shows the total pressure 
and Mach number for the case where $d_k/d_o = 0.5$, with $d_o/d_c = 0.3$ and $D_{ko}/d_o = 3$. 
The fluid solver (ANSYS-CFX) was used for this particular case.
The total pressure remains approximately constant ($\approx 1$~Torr) in the cathode channel and 
the orifice-keeper region.
Much like the case $d_o = d_k$, the supersonic expansion occurs in the keeper orifice. 
}


\clearpage

\twocolumngrid

\subsection{Keeper pressure-distance product}

We now turn our attention to the pressure-distance product in the orifice-keeper
region, a critical quantity for successful cathode ignition. 
The pressure-distance product in the keeper-orifice region is given by
\begin{equation}
    P_{ko} D_{ko} = \dfrac{P_{ko}}{P_T} \dfrac{D_{ko}}{d_o} \dfrac{d_o}{d_c} P_T d_c
    = \beta\left(\dfrac{d_k}{d_o},\dfrac{D_{ko}}{d_o},\dfrac{d_o}{d_c}\right) P_T d_c,
    \label{eqn:PkoDko-vs-PTdc}
\end{equation}
where $P_{ko}$ is the average pressure in this region, and  
$\beta$ includes both the linear and non-linear effects 
of the geometry and mass flow rate on the pressure-diameter product
with geometry. 
Linear factors and non-linear factors 
include $D_{ko}/d_o$, $d_o/d_c$, and $P_T d_c$; and $d_{k}/d_o$ (through $P_{ko}/P_T$), respectively. 

\begin{figure}
    \centering
    \includegraphics{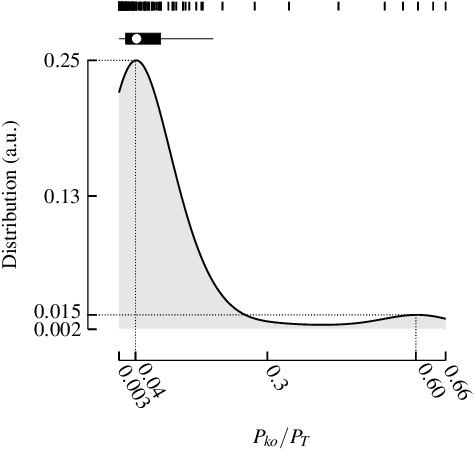}
    \caption{Combined violin and box plot of the range of keeper-to-upstream pressure ratios, $P_{ko}/P_T$.
    Each individual tick mark indicates the location of a data point.}
    \label{fig:orifice-keeper-pressure-average}
\end{figure}

We compute the keeper-orifice region pressure, $P_{ko}$, by averaging the
static pressure of all cells \textit{(i)} that are bounded by the orifice 
and keeper entrance planes and \textit{(ii)} for which the flow remains subsonic.
Those cells are part of the recirculation zone.
The range of scaled average keeper-to-upstream pressure ratios, $P_{ko}/P_T$, 
is shown in Figure~\ref{fig:orifice-keeper-pressure-average}
for the entire dataset of 87 numerical cases.
The 95\% bounds of the data are 0.004 and 0.51, respectively, with a mean of 0.04. 
This wide range of values is due to the variety of flow conditions encountered: the highest
pressure ratios ($\geq 0.35$) correspond to an entirely subsonic flow throughout the keeper-orifice region
(\textit{i.e.}, $d_k/d_o = 1$ and $P_T d_c = 1$~Torr-cm), while the lowest ones ($\leq0.01$)
correspond to a strongly underexpanded jet.

\begin{figure}
    \centering
    \includegraphics{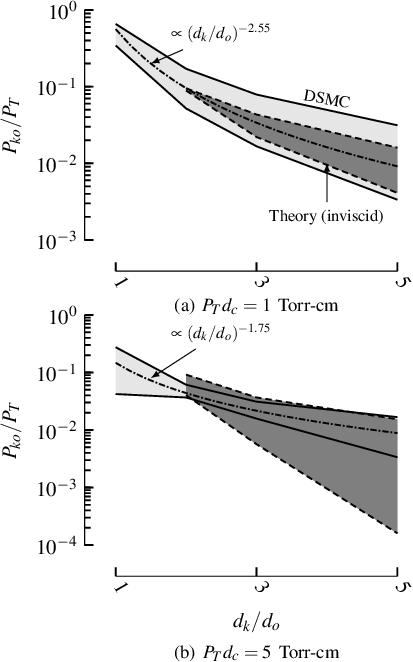}
    \caption{Keeper to upstream pressure for the entire dataset as a function of the keeper-to-orifice diameter ratio, $d_k/d_o$. $P_T d_c$ is set to (a) 1~Torr-cm and (b) 5~Torr-cm.}
    \label{fig:keeper-to-upstream-vs-dkdo}
\end{figure}

\begin{figure}[ht!]
    \centering
\includegraphics{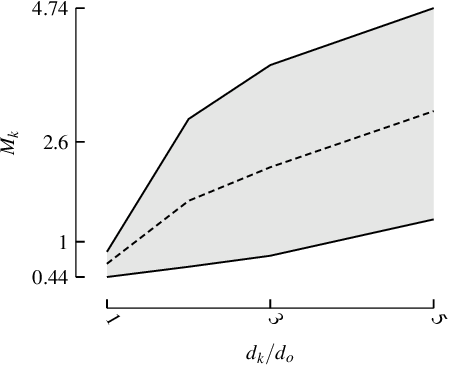}
    \caption{Keeper entrance Mach number, $M_k$, as a function of the keeper-to-orifice diameter ratio, $d_k/d_o$. The dashed line indicates the average value.}
    \label{fig:Mk-entrance}
\end{figure}

Which geometrical parameters most influence the ratio of keeper-to-upstream pressures, $P_{ko}/P_T$?
The ratio is
shown in Figure~\ref{fig:keeper-to-upstream-vs-dkdo} for
the entire dataset, as a function of the keeper-to-orifice
diameter ratio, $d_k/d_o$.
Two branches exist for $d_k / d_o = 1$: 
\begin{itemize}
    \item For $P_T d_c = 1$~Torr-cm, the flow remains subsonic in the entire
    orifice-keeper region and the keeper-to-upstream pressure ratio, is correspondingly, close to unity.
    \item For $P_T d_c = 5$~Torr-cm, the flow becomes supersonic and the
    pressure ratio subsequently decreases.
\end{itemize}
The keeper-to-upstream
pressure ratio, $P_{ko}/P_T$, varies inversely with
$d_k/d_o$:
\begin{equation}
    P_{ko}/P_T \propto \left(d_o/d_k\right)^r,
\end{equation}
where $r=2.55$ and $r=1.75$ for $P_T d_c=$ 1 and 5~Torr-cm, respectively.
A mass, momentum, and energy balance on the control volume applied to the keeper-orifice region 
shows that, under the simplifying assumption of continuum flow, 
\begin{equation}
   \dfrac{P_{ko}}{P_T} = 
   \dfrac{1}{\left(d_k /d_o\right)^2-1}
   \left[ 
    \dfrac{P_o}{P_T}
    \mathcal{F}\left(M_o, M_k\right)
    + 
    8 \dfrac{\left<\tau\right>}{P_T} \dfrac{d_k}{d_o} \dfrac{D_{ko}}{d_o}
   \right],
   \label{eqn:Pko-Pt-ratio}
\end{equation}
where $M_o$ and $M_k$ are the Mach number at the orifice exit and keeper entrance,
respectively, and $\tau$ is the axial wall shear stress. 
The function $\mathcal{F}$ is given by
\begin{equation}
\begin{split}
    \mathcal{F} = &\dfrac{M_o}{M_k}
    \left( 
    \dfrac{1+\dfrac{\gamma-1}{2} M_o^2}
    {1+\dfrac{\gamma-1}{2} M_k^2} 
    \right)^{1/2}
    \cdot 
    \left[1+\gamma M_k^2\right] \dotsm  \\
    &\dotsm -\left(1+\gamma M_o^2\right)
\end{split}
    \label{eqn:F-term}
\end{equation}
Equation~\ref{eqn:Pko-Pt-ratio} may be evaluated provided that the pressure ratio $P_o/P_T$ and the Mach number at 
the keeper entrance, $M_k$, are known. 
Because we assumed a Fanno flow in the orifice with an entrance Mach number of 0.4 and an orifice aspect ratio of 1, $P_o/P_T \approx 1/1.59$.
Figure~\ref{fig:Mk-entrance} shows
the range of numerical values of $M_k$ obtained, as a function of the keeper-to-orifice
diameter ratio. The corresponding values of $P_{ko}/P_T$ are shown
in Figure~\ref{fig:keeper-to-upstream-vs-dkdo} for the inviscid case
(\textit{i.e.}, $\left<\tau\right>=0$, for simplicity).
Reasonable agreement is obtained with the numerical results, which indicates
that the 
variation of the pressure ratio with $d_k/d_o$ can be 
explained through control-volume-based conservation laws.

\begin{figure}[ht!]
    \centering
    \includegraphics{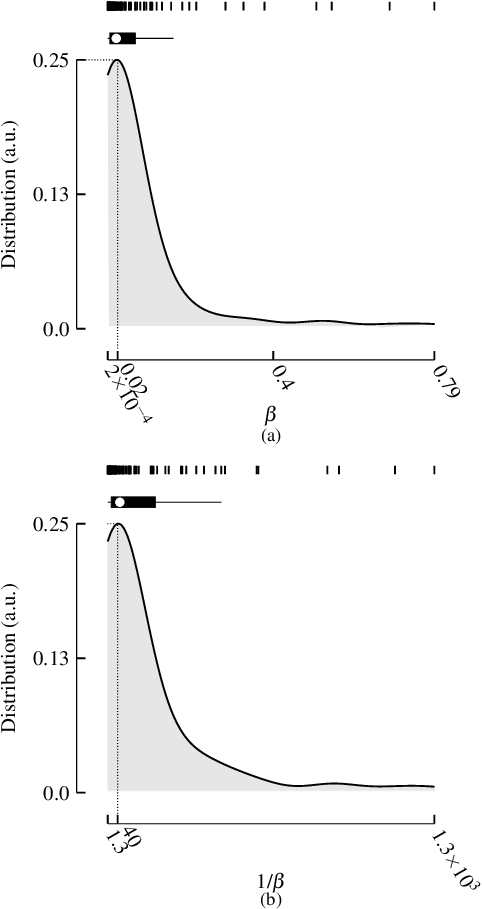}
    \caption{Combined violin and box plot of the range of values of (a) $\beta$ and (b) $1/\beta$, where $\beta = \dfrac{P_{ko}}{P_T} \dfrac{D_{ko}}{d_o} \dfrac{d_o}{d_c}$,
    for the entire dataset (including cases for which the cathode orifice is not choked).
    Each individual tick mark indicates the location of a data point.
    The data range for (b) is limited to the interval that contains 95\% of all values.}
    \label{fig:beta-plot}
\end{figure}

The distribution of the values of $\beta$ and 1/$\beta$ that we obtained for our parametric study is shown in
Figure~\ref{fig:beta-plot}. 
The 95\% interval of $\beta$ is $7\times 10^{-4}$--$3.6\times 10^{-1}$, with an average of 0.07 and a most probable value of 0.024: 
the pressure-distance product in the keeper region is 
equal to 2.4\% of the pressure-diameter product 
imposed in the upstream cathode region.

\begin{figure}
    \centering
    \includegraphics{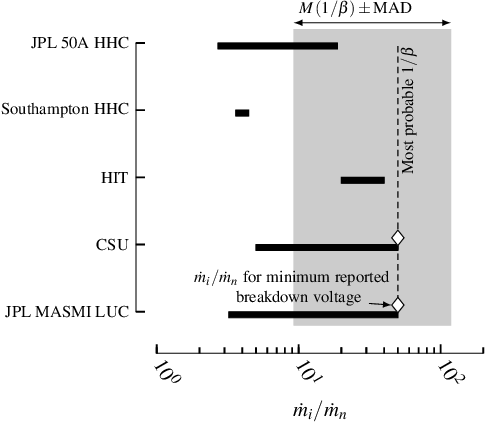}
    \caption{Ratio of ignition to nominal mass flow rates for multiple cathodes. 
    The median, MAD, and most probable value of $1/\beta$ are limited to the dataset for which $d_k / d_o > 1$ (\textit{i.e.}, the 
    cases for which the flow is choked).
    The experimental data were retrieved from Refs~\onlinecite{Payman2022,Becatti2021}
    for the HHC developed by the Jet Propulsion Laboratory (JPL)
    and from \onlinecite{Daykin2015},\onlinecite{Ning2019}, and \onlinecite{Ham2019} 
    for those developed by the University of Southampton, Harbin Institute of Technology (HIT),
    Colorado State University (CSU), respectively.
    All cathodes operate on xenon, except for the Southampton HHC, which operates on argon.
    }
    \label{fig:massflow-ignition ratio}
\end{figure}

For a choked flow, the ratio between the ignition and nominal mass flow rates, $\dot{m}_i$ and $\dot{m}_n$, is
\begin{equation}
    \dfrac{\dot{m}_i}{\dot{m}_n} = \dfrac{P_{T,i}}{P_{T,n}}\left(\dfrac{T_{T,n}}{T_{T,i}}\right)^{1/2},
\end{equation}
where $P_T$ and $T_T$ are the total, upstream conditions. The subscripts $i$ and $n$ denote the \textit{ignition} 
and \textit{nominal} conditions, respectively.
Assuming that the flow temperature remains constant ($T_{T,n} = T_{T,i}$), we have, using Equation~\ref{eqn:PkoDko-vs-PTdc}: 
\begin{equation}
    \dfrac{\dot{m}_i}{\dot{m}_n} = \dfrac{P_{T,i} d_c}{P_{T,n} d_c} = \frac{1}{\beta}\dfrac{P_{ko,i} D_{ko}}{P_{T,n} d_c} 
\end{equation}
Because most cathodes operate near $P_{T,n} d_c \sim 1$~Torr-cm\cite{Taunay2022} and with noble gases for which the ignition pressure-distance
product is near the Paschen minimum (\textit{i.e.}, $P_{ko,i} D_{ko} \sim 1$~Torr-cm), 
the last term simplifies and the ratio of ignition to nominal mass flow rates is equal to the ratio
between the upstream pressure-diameter product and the pressure-distance product: 
\begin{equation}
   \dfrac{\dot{m}_i}{\dot{m}_n} \approx \dfrac{1}{\beta} 
\end{equation}
If $d_k / d_o > 1$, this ratio has 
a median of 63.5 and a median absolute deviation of 54.2.
The most-probable value is $1/\beta \approx 50$.
Because the flow is choked, this suggests that a fifty-fold increase 
in mass flow rate results in a favorable pressure-distance product for a Paschen-like discharge.
Increasing the mass flow rate prior to cathode ignition is a strategy that is typically used in orificed-emitter hollow cathodes.\cite{Becatti2021,Ham2019}
We show in Figure~\ref{fig:massflow-ignition ratio} the ratio of experimentally observed 
ignition flow rate to nominal flow rate for a few cathodes operating on both argon and xenon, along with
the range of favorable flow rate ratios as determined from the median, $M$, median absolute deviation, MAD,
and most-probable value.

\section{Conclusion}
We have performed a parametric study to quantify the effect of the changes in keeper geometrical parameters on the 
neutral flow field within hollow cathodes. 
Because cathode ignition occurs typically between the keeper and orifice plates, we focused our study on the keeper-orifice region.
Using the DSMC method, we showed that the neutral cathode flow results in most cases in an underexpanded, 
supersonic, rarefied jet with Knudsen numbers above 0.1, along with a toroidal subsonic recirculation zone.
The recirculation zone is bounded by the underexpanded jet viscous layer, the cathode orifice plate, and the keeper plate.
We have shown that simple conservation laws can explain the experimentally and numerically observed variation of the ratio between 
the keeper-orifice region pressure and the total, upstream pressure.
Good agreement is obtained with our control-volume based approach and numerical results.
The statistical analysis we conducted on the numerical data demonstrated that \textit{(i)} 
the pressure within the keeper-orifice region is most likely
4\% of the total, upstream pressure, although cases for which the flow remains subsonic within the keeper-orifice region result in
much higher pressure recovery values (up to 80\%), and
\textit{(ii)} $1/\beta$, the ratio between the cathode pressure-diameter product, $P_T d_c$, and the keeper-orifice region pressure-distance ratio, $P_{ko} D_{ko}$, is most likely $\approx 50$ (excluding the case where $d_k = d_o$) and is related to the ratio of ignition to nominal mass flow rates. 

Our study is limited to that of the neutral {xenon} flow in cathodes and 
does not consider any plasma effects (\textit{e.g.}, neutral heating due 
to charge-exchange collisions) that may alter the flow structure.
However, the neutral flow field obtained from DSMC may be used as a first guess
in a coupled plasma / neutral numerical model (\textit{e.g.}, PIC/DSMC).
The study is further limited by the data upon which it is based:
we considered a wide, albeit non-exhaustive, range of rescaled parameters describing orificed cathodes with an enclosed keeper, and
our conclusions may not apply to cathode configurations that are not covered by our dataset.

The study demonstrates that the flow within the orifice, prior to cathode ignition, is more accurately represented by a Fanno flow approach than
an isentropic flow or Poiseuille flow one. 
The Poiseuille flow model should not be used to estimate the pressure (and/or density and temperature) within hollow cathodes
because most assumptions upon which the it is based are invalid in the regime in which cathodes operate.
Select numerical cases also show that fluid solvers, while offering qualitative agreement with DSMC results in the keeper-orifice region, differ quantitatively from DSMC in regions where the gas becomes transitional
and rarefied.
Finally, as shown through our statistical analysis and explained by simple conservation laws, 
a fifty-fold increase in mass flow rate is likely to result in a minimum voltage in the Paschen-like, DC discharge.
Although we only considered xenon, these conclusions are likely applicable to other noble, monatomic gases such as argon:
{while a lighter gas will have a lower mass flow rate for the same total pressure, temperature, and geometry (see Equation~\ref{eqn:choked-mass-flow-rate}),
all other scaled parameters (Mach number, Knudsen number, pressure-diameter product, 
ratio of specific heats, geometric parameters) remain similar.}

\section*{Acknowledgments}
{
The views expressed in this article are those of the authors and do not reflect the official policy or position of the U.S. Naval Academy, Department of the Navy, the Department of Defense, the U.S. Government, or any agency thereof.}
This work was conducted at the Princeton Collaborative Research Facility supported by the U.S. Department of Energy through contract DE-AC02-09CH11466.
The computational resources of the United States Naval Academy are gratefully
acknowledged.

\section*{Author declaration}
\subsection*{Data availability}
The data that support the findings of this study are available from the corresponding author
upon reasonable request.

\subsection*{Conflict of interest}
The authors have no conflicts to disclose.

\section*{Appendix}
\setcounter{equation}{0}

\subsection{Cathode dimensions}

The dimensions of the considered cathodes are shown
in Table~\ref{tbl:allcathodes-dimensions}. 
The Southampton and Harbin Institute of Technology cathodes, 
MasMI LUC, JPL 50~A HHC are heaterless hollow cathodes. 
The corresponding range of normalized parameters is:
\begin{itemize}
    \item $L_o/d_o$: 1.1--1.5
    \item $d_o/d_c$: 0.1--0.8 (excluding tube cathodes)
    \item $D_{ko}/d_o$: 0.31--28
    \item $d_k/d_o$: 1--7
\end{itemize}

\begin{table*}[t]
\begin{ruledtabular}
     \centering
    \begin{tabular}{cccccccc}
Name &  Insert & \multicolumn{2}{c}{Orifice}           & \multicolumn{3}{c}{Keeper} & Refs. \\
     & diameter & diameter & length & orifice diam. & distance & thickness &      \\ 
     & $d_c$ & $d_o$ & $L_o$ & $d_k$ & $D_{ko}$ & $t_k$ & \\
Siegfried & 3.9 & 0.5--1 & --- & 3.6 & 2.5 & & \onlinecite{Siegfried1982} \\
NEXIS & 12.0 & 2.75 & 1.75 & 4.8 & 1, 2.4 & 1.3 & \onlinecite{Polk1999,Mikellides2004}\\
NSTAR & 3.8 & 1.02 & 1.52 &	4.65 & 1 & 1.5 & \onlinecite{Goebel2007,Polk1999}\\
JPL 1.5~cm & 7.0 & 3--5 & --- & 6.4 & --- & --- & \onlinecite{Goebel2007LaB6} \\
Southampton HHC & 2 & 0.75--2 & --- & 2--7.5 & 1--21 & 2 & \onlinecite{Daykin2015}\\
MasMI LUC & 1.82 & 0.7 & 0.8 & 1.5 & 0.4 & 3 & \onlinecite{Becatti2021}\\
JPL 50 A HHC & 6.3 & 4.8 & --- & 5.7 & 1.5 & --- & \onlinecite{Payman2022} \\
Harbin Institute & 2 & 0.8 & 1.5 & 2.5 & 1.5 & 2 & \onlinecite{Ning2019}
    \end{tabular}
    \caption{Dimensions of orificed hollow cathodes with enclosed keeper from the literature. Values are in mm.}
    \label{tbl:allcathodes-dimensions}
\end{ruledtabular}
\end{table*}

\renewcommand{\theequation}{B\arabic{equation}}
\subsection{Keeper-orifice region conservation laws}

\begin{figure*}[t]
    \centering
    \includegraphics{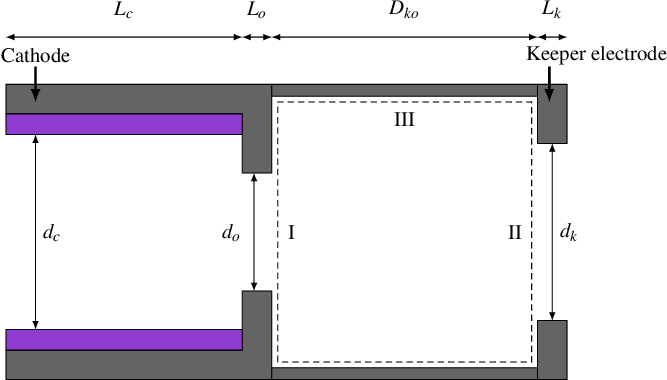}    
    \caption{Control volume considered.}
    \label{fig:control-volume-derivation}
\end{figure*}

We derive here Equation~\ref{eqn:Pko-Pt-ratio} following a framework similar to that of Emmert~\textit{et al.}\cite{Emmert2009}
The control volume considered that is comprised of surfaces I, II, and III is shown in Figure~\ref{fig:control-volume-derivation}.
We assume that:
\begin{enumerate}
    \item The flow is in the continuum regime
    \item The flow is adiabatic 
    \item The static pressure on the walls adjacent to surfaces I, II, and III is constant and equal to the keeper-orifice region pressure, $P_{ko}$. 
\end{enumerate}
The subscripts $o$ and $k$ will be used to denote the quantities defined at the 
cathode orifice and keeper entrance surfaces, respectively.

\subsubsection{Continuity}

From the conservation of mass applied to the CV,
\begin{equation*}
    \rho_o u_o \pi r_o^2 = \rho_k u_k \pi r_k^2. 
\end{equation*}
Using the perfect gas law, $P = \rho R_g T$, and the definition of the
Mach number, $M = u/\sqrt{\gamma R_g T}$, we obtain
\begin{equation}
   \dfrac{P_o}{P_k} = \dfrac{M_k}{M_o}\sqrt{\dfrac{T_o}{T_k}} \bar{d}^2, 
   \label{eqn:continuity-derivation}
\end{equation}
where $\bar{d} = d_k/d_o$ and all thermodynamic quantities are static (as opposed to stagnation) quantities.

\subsubsection{Energy}

Because the flow is adiabatic, conservation of energy applied to the CV yields
\begin{equation*}
    h_{T,o} = h_{T,k},
\end{equation*}
where $h_{T}$ is the total enthalpy.
Using the definition of the total enthalpy,
\begin{equation*}
    h_T = \dfrac{\gamma}{\gamma-1} R_g \left[ 1+\dfrac{\gamma-1}{2} M^2\right],
\end{equation*}
we obtain
\begin{equation}
    \dfrac{T_o}{T_k} = \dfrac{1+\dfrac{\gamma-1}{2} M_k^2}{1+\dfrac{\gamma-1}{2} M_o^2}.
    \label{eqn:energy-derivation}
\end{equation}

\subsubsection{Momentum along $z$-axis}

The conservation of momentum along the $z$-axis is 
\begin{equation}
\begin{split}
-\left[ \left(\rho_o u_o^2 + P_o\right) \pi r_o^2 + P_{ko} \pi \left(r_e^2-r_o^2\right)\right] \\
    +\left[ \left(\rho_k u_k^2 + P_k\right) \pi r_k^2 + P_{ko} \pi \left(r_e^2-r_k^2\right)\right]& \\
    + \Pi_{rz} 2\pi D_{ko} r_e &= 0.
\end{split}
\end{equation}
Using the definition of the Mach number, we obtain, after multiplying by $1/\left(P_T r_o^2\right)$ and simplifications:
\begin{equation}
\begin{split}
    \dfrac{P_{ko}}{P_T} \left(\bar{d}^2-1\right) = 
    \dfrac{P_k}{P_o}\cdot\dfrac{P_o}{P_T} \bar{d}^2\left(1+\gamma M_k^2\right) & \\
    -\dfrac{P_o}{P_T} \bar{d}^2\left(1+\gamma M_o^2\right )&
    + \dfrac{\left<\tau\right> 2 D_{ko} r_e}{P_T ro^2}.
\end{split}
    \label{eqn:momentum-step2}
\end{equation}
Inserting Equations~\ref{eqn:continuity-derivation} and~\ref{eqn:energy-derivation} into Equation~\ref{eqn:momentum-step2},
and, because $r_e \approx 2 r_k$, we have
\begin{equation}
    \dfrac{P_{ko}}{P_T} \left(\bar{d}^2-1\right) = 
    \dfrac{P_o}{P_T} \mathcal{F}\left(M_o,M_k\right)
    + 8 \dfrac{\left<\tau\right>}{P_T} \bar{d} \dfrac{D_{ko}}{d_o},
\end{equation}
where $\mathcal{F}$ is given in Equation~\ref{eqn:F-term}.

\clearpage
\section*{References}

\bibliography{biblio}

\end{document}